\documentclass[12pt]{iopart}

\usepackage{graphicx}
\usepackage{iopams}
\usepackage{subfigure}
\usepackage{color}
\usepackage{hyperref}
\usepackage{comment}

\definecolor{darkblue}{rgb}{0,0,.7}
\definecolor{darkred}{rgb}{0.7,0,0}

\hypersetup{colorlinks=true, linkcolor=darkblue,citecolor=darkblue,urlcolor=darkblue}

\begin{document}

%\title[Electromagnetic design]{Electromagnetic design of the KATRIN large-volume air coil system}
\title{Electromagnetic design of the KATRIN large-volume air coil system}

\author{Ferenc  Gl\"uck$^1$, Guido Drexlin$^1$, Benjamin Leiber$^1$, Susanne Mertens$^{1,2}$, 
Alexander Osipowicz$^3$, Jan Reich$^1$ and Nancy Wandkowsky$^1$}

\address{$^1$ KCETA, Karlsruhe Institute of Technology, Karlsruhe, Germany}

\address{$^2$Lawrence Berkeley National Laboratory, Berkeley, USA}

\address{$^3$University of Applied Sciences, Fulda, Germany}

\ead{ferenc.glueck@kit.edu}

\begin{abstract}
 
The KATRIN experiment is designed to determine the absolute neutrino mass scale 
with a sensitivity of 200 meV (90 \% CL)
by measuring the  electron energy spectrum close to the endpoint of  molecular tritium $\beta$ decay.
Electrons from a high-intensity gaseous tritium source
are guided by a strong magnetic field of a few T
to the analyzing plane of the main spectrometer where
an integral energy analysis takes place in a low field region (B$<$0.5 mT).
An essential design feature to obtain adiabatic electron transport through 
this spectrometer is a large volume air coil system surrounding the vessel. The system has two key tasks:
to adjust and fine-tune
 the magnetic guiding field (Low Field Correction System), as well as to compensate the distorting effects of the earth
magnetic field (Earth Field Compensation System). In this paper we outline  the key electromagnetic design issues 
for this very large air coil system, which allows for well-defined electron transmission and optimized background
reduction in the KATRIN main spectrometer.

\end{abstract}

%\maketitle

%\tableofcontents

\section{Introduction}
 \label{SecIntro}

Experimental information about the neutrino masses and lepton mixing 
is important both for particle physics and  cosmology.
The observation of flavor oscillations of atmospheric, solar, reactor and accelerator neutrinos has provided convincing evidence for lepton 
mixing and non-zero neutrino masses. 
However, neutrino oscillation studies  only allow to access the mass splittings of various neutrino 
mass eigenstates, but yield no information on the absolute neutrino mass scale.

Cosmological observations \cite{Pastor}
and neutrinoless double beta decay experiments \cite{Giuliani}
provide access to the absolute neutrino mass scale,
but  are rather model-dependent. On the other hand, 
a direct and model-independent way to measure the effective electron neutrino mass
is possible by high-precision $\beta$-spectroscopy of nuclear $\beta$-decays
close to the endpoint. 
The $\beta$-emitter with the best decay characteristics ($t_{1/2}=12.3$ y and end point energy
$E_0=$18.6 keV) is tritium  \cite{OttenWeinheimer,Drexlin}.

\begin{figure}[htbp]
\centering
\includegraphics[width=0.80\textwidth]{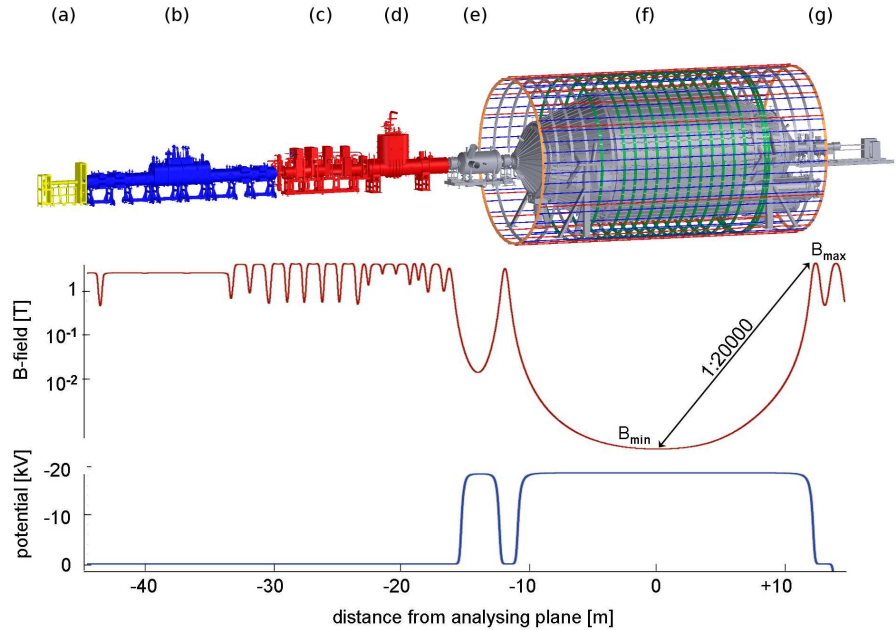}
\caption{The KATRIN experimental setup with its main components: a, rear section; b, windowless gaseous tritium source
(WGTS); 
c, differential pumping section (DPS); d, cryogenic pumping
section (CPS); e, pre-spectrometer; f, main spectrometer with large-volume air coil system;
 g, focal plane detector. Below, the magnetic field and the electric
 potential along the beam axis are displayed.}
\label{FigBeamline}
\end{figure}

The Karlsruhe Tritium Neutrino (KATRIN) experiment \cite{KATRIN} is designed to determine the absolute neutrino mass scale
with a sensitivity of 200 meV
by a precise measurement of the
electron energy spectrum close to the endpoint $E_0$ of molecular tritium.
In the 70 m long setup (see Fig. \ref{FigBeamline}), 
electrons are guided from the source  to the detector by magnetic fields in the range of 
 a few T, which are created by many superconducting coils.
The main spectrometer of the MAC-E filter type is on high negative potential (around -18.6 kV) and acts  
as an electrostatic filter for the
integral energy spectrum measurement. In this filter type,
only electrons with  enough kinetic energy are able to be transmitted through the
spectrometer to be counted at the detector.
Inside the main spectrometer, we need a small magnetic field
(below 0.5 mT),  to convert most of the transversal energy of the $\beta$-decay electrons into longitudinal energy
by the inverse magnetic mirror effect.
To fine-tune this  magnetic field for the purposes of the precise energy filtering 
and to compensate the disturbing effect of the
earth magnetic field, a large volume
(about 3000 ${\rm m}^3$) air coil system has been designed and built.

The main purpose of this paper is to discuss the most important electromagnetic
design features of this coil system. 
The technical design of the system and results of corresponding magnetic field measurements 
will be presented in a second publication \cite{Reich2013}.

The plan of this paper is the following.
In Sec. \ref{SecKatrin} we give a short overview of the main KATRIN components, and
point out the key design requirements
that are relevant for the successful air coil operation.
In  Sec. \ref{SecAdiabtrans} we discuss the adiabatic longitudinal and transmission energy of electrons  and
also define the notion of analyzing point and the transmission condition. Then
we explain why and how the transmission condition in the main spectrometer should be fulfilled.
Sec. \ref{SecRequirements} contains a description of the most important requirements about
the magnetic field inside the main spectrometer, and the specific role of the air coil system to fulfill
these requirements is explained.
Sec. \ref{SecLFCS} is devoted to a detailed explanation of the axisymmetric part of the air coil system
(LFCS), and in Sec. \ref{SecEMCS} the non-axisymmetric earth field compensating part (EMCS)
is described. In  \ref{appendix-fieldsim} we give
a short overview about the magnetic and electric field simulation methods that have been used for the
air coil design.  Finally, in \ref{appendix-mathoptim} we
present a multiobjective mathematical optimization method
that is useful to compute various LFCS coil current configurations.

\section{The KATRIN experiment}
 \label{SecKatrin}

In this section we give a short overview of the main components of the KATRIN experiment, emphasizing those details
that are important for the electromagnetic design issues of the KATRIN air coil system.
For more details about the KATRIN experiment we refer to Refs. \cite{OttenWeinheimer,Drexlin,KATRIN,Wolf,Thummler}.

The 70 m long KATRIN setup (see Fig. \ref{FigBeamline}) contains the following main components:

\vspace{2mm}
{\bf WGTS}
\vspace{2mm}

High-purity molecular tritium gas  with a temperature of 30 K is injected into the middle of the 10 m long and
9 cm diameter tube of the WGTS (Windowless Gaseous
Tritium Source). The injected gas  diffuses to both ends of the WGTS beam tube, where it is
 pumped out to  a large degree by a total of 4 pumping ports \cite{WGTSMonitoring}. A system of
21 superconducting coils generates a
high (3.6 T -- 5.6 T) magnetic  field, which guides the $\beta$-decay electrons out of the
source along  magnetic field lines.

\vspace{2mm}
{\bf DPS and CPS}
\vspace{2mm}

The transport section downstream of the WGTS consists of 
two main tritium retention systems:
 the DPS (Differential Pumping Section) and the CPS
(Cryogenic Pumping Section). Both components together
eliminate the remaining tritium gas from the 
beamline, thus preventing tritium migration to
the main spectrometer. This is of major importance, as
even trace amounts of tritium  in the main spectrometer would
cause  an untolerably large
background rate and initiate large systematic effects, through the $\beta$-decays of the tritium molecules.
To prevent this,
the tritium gas at first is differentially pumped
out at the 4  main pump ports of the DPS \cite{Lukic}. 
Second, the remaining tritium is trapped
on to the cold inner surfaces of the CPS \cite{Gil}. 
The beam tube of both cryostats
is operated at high magnetic fields up to 5.5 T, in order to
guide the $\beta$-decay electrons  towards the main spectrometer.
This magnetic field is created by 5 and 7 superconducting coils in the DPS and CPS, respectively.
Some of these coils are not coaxial with the main beamline, in order to reduce the 
molecular beaming effect \cite{Luo}.
As the transport section also filters out positive ions, only electrons are transmitted to the electrostatic spectrometers
for energy analysis.

\vspace{2mm}
{\bf Pre-spectrometer}
\vspace{2mm}

At first, a
smaller spectrometer (pre-spectrometer)
at the entry of the spectrometer section
allows to filter out the low-energy part of the $\beta$-spectrum
which is not important for the neutrino mass determination (since the energy is not close to the endpoint).
In fact, the potential of the pre-spectrometer can be adjusted from 0 up to -18.3 kV, thus optimizing the
background level as a function of the filter potential \cite{Prall}.
The pre-spectrometer has two superconducting coils at the ends: both of them have a reference field of  4.5 T
at the coil centre, and they  generate a 15 mT field in the middle of this spectrometer.
In the following 
we refer to the source-side magnet (between CPS and pre-spectrometer) as PS 1 coil, and the other one (between
pre- and main spectrometer) as PS 2 coil.
Due to the magnetic field of these coils, 
the $\beta$ electrons are adiabatically guided through the pre-spectrometer,
even when operated at low or zero potential \cite{Prall}.

\vspace{2mm}
{\bf Main spectrometer}
\vspace{2mm}

The very large main spectrometer (length 23.6 m, diameter 10 m) has the task of precision energy filtering so that
only electrons with  high enough kinetic energy are able to overcome the electrostatic retarding potential to be
transmitted to  the detector for counting.
All electrons with smaller kinetic energy are reflected and move back to the source.
However, the electric field inside the main spectrometer is able to filter only the longitudinal kinetic energy 
$E_\parallel$ of the electrons,
but not the transversal energy $E_\perp$ 
(the longitudinal energy is defined by the electron velocity component parallel to the magnetic
field direction). As $\beta$-decay electrons in the source are created with isotropic angular distribution,  
a significant part of their energy can be transversal. 
If their transversal energy component $E_\perp$ remained unaltered,
most of the electrons with total
energy near the endpoint would not reach the detector, 
resulting in a rather poor statistics. The solution for this problem
is to significantly reduce the magnetic field strength towards the center of main spectrometer. 
The corresponding field configuration  has been designed  to first order so that the motion
 of $\beta$-electrons  in the KATRIN system is adiabatic \cite{Jackson,Prall}. Therefore
 the first adiabatic invariant (proportional to transversal energy per magnetic field)
is approximately constant (see Eq.    \ref{gammamu} in the next section for the relativistic expression of this adiabatic invariant).
Consequently, when the $\beta$-decay electrons move from high to small magnetic field
(i.e. from the entry to the centre of the main spectrometer),
most of their transversal kinetic energy is converted into longitudinal energy.
In doing so, 
it is important to keep  the appropriate order: 
first the conversion of  transversal to longitudinal  energy  has to take place
before the reflecting electric field  `eats up` all the longitudinal energy  of the electron (see the next section for more details).

Due to the non-zero magnetic field inside the main spectrometer,
this  conversion is not perfect, thus the electrons will retain 
a small transversal energy.
As this energy is not scanned by the electrostatic retarding potential, it
also defines the energy resolution of the experiment. 
With a reference value of the magnetic field in the middle of the main
spectrometer of  0.3 mT (which is 20000 times smaller than the maximal field of 6T in the KATRIN setup),
the energy resolution of KATRIN (defining the width of transmission from $0\to 100$ \% for an isotropic source)
will be 0.93 eV at 18.6 keV electron energy.

The conversion from transveral to longitudinal energy is also called magnetic adiabatic collimation (the electron velocity
directions are collimated parallel to the magnetic field), and a spectrometer using electrostatic retardation
together with magnetic adiabatic collimation, like the 
KATRIN main spectrometer, is called a MAC-E filter \cite{LobashevSpivak,Picard}.
Thus the KATRIN experiment, like the pioneering
Mainz \cite{Mainz} and Troitsk \cite{Troitsk}  neutrino mass experiments, will make use of
 the MAC-E filter principle to measure the neutrino mass.

A $\beta$-decay electron coming from the source follows a specific  magnetic field line, to a good approximation.
Therefore the $\beta$-decay electrons created inside the transported
magnetic flux tube (defined by the reference magnetic flux value of 191 T${\rm cm}^2$) will always
remain inside this flux tube until they are counted
by the detector. Since the magnetic field in the main spectrometer will  be a factor of $10^4$  times smaller than 
the field $B_s$ in the source,
the diameter of the flux tube has to be enlarged by a factor of
100 relative to the source.
Therefore, the main spectrometer diameter has to be very large (about 10 m).
In order to minimize electron interactions with residual gas molecules, the main spectrometer
should also feature an excellent ultrahigh vacuum.

The main spectrometer has 3 nearby superconducting coils: at the source side the abovementioned PS 2 coil,
and at the detector side the pinch (PCH) and the detector (DET) coils.
The latter two together have a significantly larger magnetic moment than the PS 2 coil alone, therefore
the magnetic field of the superconducting coils inside the main spectrometer is asymmetric: it is larger at the detector side
than near the source side (note that the stray field of a coil is proportional to its magnetic moment).

Table \ref{TabSuper} shows the central axial positions and the typical maximal fields of the 
3 superconducting coil systems (WGTS, DPS, CPS) and the 4 superconducting coils
(PS1, PS2, PCH, DET).
In addition, this table presents the contributions of the various superconducting coil systems
to the magnetic field at the center ($z=0$) of the main spectrometer. In Sec. \ref{SecLFCS} we explain
the negative sign of  these field values.

\begin{table}
\begin{center}
\vspace{2mm}
\begin{tabular}{| l   |  c   |  c   |   c  |} \hline
 Field source        & $z_c$ (m) & $B_c$ (T) &  $B_{z0}$ ($\mu$T)   \\ 
\hline \hline
Earth            & - & - &  20  \\
\hline
WGTS coil system & -38.87 & 3.6 &  -9.7 \\
\hline
DPS coil system & -27.25 & 5 & -16.3  \\
\hline
CPS coil system & -20.58  & 5.6 & -38.2 \\
\hline
PS 1  coil   &  -16.46 & 4.5 & -18.5  \\ 
\hline
PS 2 coil    &  -12.10 & 4.5 & -46.5  \\ 
\hline
PCH (pinch) coil    &  12.18 & 6 & -65.2  \\ 
\hline
DET (detector) coil    &  13.78 & 3.6 & -48.4  \\ 
\hline
\end{tabular}
\caption{Axial magnetic field contributions $B_{z0}$ 
at the center of main spectrometer ($z=r=0$)
from the horizontal earth field and from the various superconducting coil systems and coils. 
$z_c$ is the central axial position of the coil system, and $B_c$ is the typical maximal field
near this position.
\label{TabSuper}} 
\end{center}
\end{table}

Besides the fields of the s.c. coils, there is a  non-negligible contribution from the earth magnetic field
whose  vertical and horizontal components at the location of the KATRIN experiment are  43.6 
$\mu$T  and 20.6 $\mu$T, respectively \cite{Reich2009,Macmillan,Online}.
The 20 $\mu$T earth field value 
in table \ref{TabSuper} represents 
 that component of the horizontal earth magnetic field which is
parallel to the  spectrometer axis; the horizontal perpendicular earth field component 
is 5 $\mu$T. These values result from the fact that the KATRIN beamline is aligned almost to south-north direction,
with an angle of $14^\circ$ relative to the horizontal earth field.

In addition to the coils and the earth field, the magnetic field in the main spectrometer can be
distorted by magnetic materials in the spectrometer building surrounding the spectrometer vessel.
In particular, parts of the concrete reinforcements in the building contain normal steel.
In this context it should be emphasized that extensive careful design works 
were performed, prior to construction of the building,
to reduce these effects by employing stainless steel reinforcements
(mainly below the spectrometer vessel), to minimize the influence of magnetic field disturbances due to
normal steel  \cite{Gluck2005a}. Extensive field measurements inside the spectrometer tank
\cite{Reich2009,Reich2012} have revealed the success of these measures, as 
the magnetic field in the middle plane of the tank due to the remanent magnetization of the magnetic materials 
is smaller than 2 $\mu$T.

The whole inner surface (700 ${\rm m}^2$) of the main spectrometer tank is covered by a wire electrode system 
 to reduce the background due to secondary electrons coming from cosmic muon interactions in the vessel hull,
and also to refine and stabilize the electric field inside the tank. This wire system consists of 240 wire modules, 
with a  total wire length of 42 km \cite{Valerius}. 
Most of the wire modules have a double wire layer, and only the smaller wire module rings at the entrance and
exit regions of the main spectrometer tank (at the steep cone)
have a single layer. In the standard electric potential mode the outer and inner wire
layers will
be on a potential which is
100 V and 200 V more negative than the tank, respectively.
Accordingly, the single layer modules will be 100-250 V more positive than the inner
wires, in order to fulfill the transmission condition (see sections \ref{SecAdiabtrans} and \ref{SecLFCS} for more details).

\vspace{2mm}
{\bf Detector}
\vspace{2mm}

The transmitted electrons are counted by a segmented silicon PIN-diode detector
with 148 pixels, which is located 
inside the warm bore of the detector magnet DET \cite{Harms,Amsbaugh}.
The energy resolution  of the detector is better than 1.5 keV (FWHM),
which is sufficient to discriminate signal electrons from continuum background.
It is possible to elevate the detector on positive potential (up to 10 kV at present), in order
to shift signal electrons into a favorable region-of-interest.
The standard
central field of the detector coil without using this post-acceleration option is 3.6 T (the value we have used for the simulations
in this paper).  If the post-acceleration is turned on, one can increase the detector coil field up to 6 T.

\section{Adiabatic transmission}
 \label{SecAdiabtrans}

The motion of electrons with small transversal energy in the KATRIN main spectrometer is approximately adiabatic
(see Ref. \cite{Prall}, sec. 8).  Thus they follow  the magnetic field lines to very good approximation
(apart from a small magnetron drift
perpendicularly to the field lines). In addition, the first adiabatic invariant
\begin{equation}
\label{gammamu}
\gamma\mu=\frac{\gamma+1}{2} \frac{E_\perp}{B}
\end{equation}
is constant during the motion. Here $B$ denotes the magnetic field, $E_\perp$ the transversal kinetic energy, 
$\gamma=1/\sqrt{1-v^2/c^2}=1+E/(mc^2)$ 
the  relativistic Lorentz factor (with electron mass $m$ and kinetic energy $E$),
while $\mu$ denotes the orbital magnetic moment of the electron (see sec. 12.5 of \cite{Jackson}).
For the following discussion of electron transmission condition through the main spectrometer,
let us consider an electron starting at point ${\bf P_s}$ in the source
with kinetic energy $E_s$ and polar angle $\theta_s$ 
between velocity direction 
and magnetic field. The electric potential and magnetic field at this point will be denoted by $U_s$ and $B_s$, respectively. 
The kinetic energy
$E$ of the electron at an arbitrary point ${\bf P}$ along its trajectory can then be calculated from energy conservation:
$E_s-eU_s=E-eU$, where $U$ is the electric potential at point ${\bf P}$, and $e$ denotes the unsigned electron 
charge ($e>0$).
The adiabatic longitudinal energy at point ${\bf P}$ is then:
\begin{equation}
\label{Elong}
E_\parallel=E_s+e(U-U_s)-\frac{B}{B_s} \frac{\gamma_s+1}{\gamma+1}E_s \sin^2\theta_s,
\end{equation}
with magnetic field $B$ at point ${\bf P}$ and the relativistic factors  $\gamma$ and $\gamma_s$ at points ${\bf P}$ and
${\bf P_s}$, respectively.

Let us first consider only small  starting angles  so that $(B/B_s) \sin^2\theta_s<1$ is fulfilled everywhere
between source and detector (absence of magnetic mirror reflection). In this case,
for large enough starting energy $E_s$, the adiabatic longitudinal energy is  positive everywhere along the electron trajectory. 
This means
that the electron is transmitted, i.e. it reaches the detector (assuming adiabaticity). 
Now, we define the analyzing point ${\bf P_A}$ as the point along the magnetic field line where the longitudinal
energy has its minimal value. Decreasing the starting kinetic energy $E_s$, there exists a 
transmission energy $E_s=E_{\rm tr}$ so that the longitudinal energy is zero at the analyzing point
${\bf P_A}$, while at other points still being positive. This transmission energy has the expression
\begin{equation}
\label{Etransmission}
%E_{\rm tr}=\frac{e(U_s-U_A)}{1-\frac{B_A}{B_s} \frac{\gamma_s+1}{\gamma_A+1} \sin^2\theta_s},
E_{\rm tr}=\frac{e(U_s-U_A)}{1-(B_A/B_s) \left[(\gamma_s+1)/(\gamma_A+1) \right]\sin^2\theta_s},
\end{equation}
where $U_A$ and $B_A$ denote the electric potential and magnetic field at the analyzing point. It is obvious from the above
definition that $E_{\rm tr}$ corresponds to a transmission limit: for starting energies above the transmission energy
($E_s>E_{\rm tr}$) the electron is transmitted  and reaches the detector, while
 for energies below this limit ($E_s<E_{\rm tr}$) the electron  is reflected back towards the starting point 
 and does not get to the detector. 
With this definition it is evident that
the reflection can occur only before
or at the point ${\bf P_A}$,
so once the electron propagates to ${\bf P_A}$ it will reach the detector.
In order to compute the adiabatic transmission energy, the electric potential and magnetic
field values at the starting point ($U_s$ and $B_s$) and at the analyzing point  ($U_A$ and $B_A$)
have to be known.
Note that, generally, the analyzing point ${\bf P_A}$ and therefore also 
the corresponding $U_A$ and $B_A$ values depend on the starting polar angle  $\theta_s$.

The knowledge of the above transmission energy is crucial in order to compute the transmission function, 
which is the probability that an electron with  fixed starting energy is transmitted. 
The transmission function depends explicitly on the starting energy;
in the adiabatic approximation it is an increasing function of $E_s$ (in some regions of
$E_s$ it is constant). 
Importantly, it depends strongly on the starting angle distribution of the electrons
(through the $\theta_s$ dependence of $E_{\rm tr}$; see Eq. \ref{Etransmission}).
To calculate the transmission function, first one has to find (for a given starting energy $E_s$)
the angular transmission region, i.e. those values of $\theta_s$ for which
$E_{\rm tr}<E_s$ is fulfilled. Second, one has to
integrate the normalized electron angular distribution over this region. Due to the $\theta_s$ dependence
of $U_A$ and $B_A$ (in the general case) this calculation can be rather complicated.

For zero starting angle ($\theta_s=0$), the analyzing point  is 
where the absolute potential $|U|$ attains its maximal value
(let us denote this point by ${\bf P_A^0}$). 
In the following we assume that the main spectrometer electrode system
displays a mirror symmetry relative to the center ($z=0$) of the spectrometer vessel.
 In this case, for the on-axis field line ($r=0$) the point ${\bf P_A^0}$ is at $z=0$.
For off-axis field lines the axial coordinate of the point ${\bf P_A^0}$ can be different from zero.
However, it is zero if the field line is symmetric to  the $z=0$ plane
(in that case  the radial component of the magnetic field at $z=0$ vanishes).

\begin{figure}[htbp]
\centering
\includegraphics[width=0.50\textwidth]{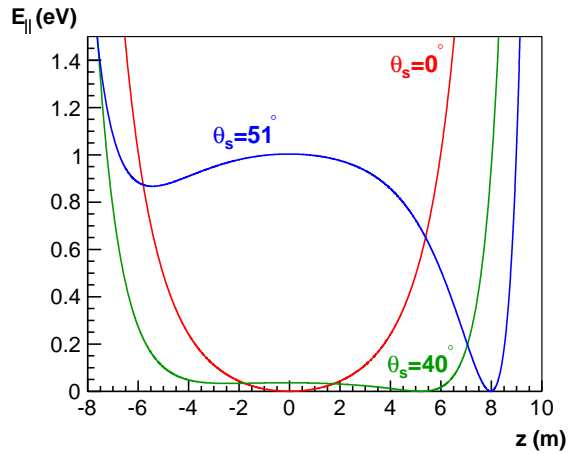}
\caption{Example for violation of the adiabatic transmission condition  for an on-axis field line.
The curves display the
 longitudinal energy  for 3 different
starting angle values.}
\label{FigLong0}
\end{figure}

On the other hand, for finite starting angles the magnetic field can shift the point  ${\bf P_A}$ away from ${\bf P_A^0}$:
in this case, the analyzing point  ${\bf P_A}$ depends on the starting angle $\theta_s$. 
This can happen if the electric potential is rather homogeneous close to the point 
${\bf P_A^0}$, and the magnetic field has a minimum value at ${\bf P_A^0}$ and is 
rather inhomogeneous near this point. Namely, in this case the third magnetic field term in eq. \ref{Elong} decreases the longitudinal
energy when the point moves away from ${\bf P_A^0}$ (due to the increasing magnetic field), 
while the slow increase of the $eU$ term
is not able to compensate this decrease. Therefore, the longitudinal energy minimum will not be at ${\bf P_A^0}$, but somewhere
farther away, where the electric potential becomes more inhomogeneous. In this case,
the analyzing point ${\bf P_A}$ and thus the $U_A$ and $B_A$ values depend on  $\theta_s$, 
and a rather complicated procedure
is required to determine the transmission function.
In this case we say that the transmission condition --- i.e. the independence of the analyzing point from the
starting angle --- is not fulfilled.
Fig. \ref{FigLong0} shows an example of the behaviour of the longitudinal energy 
$E_\parallel$ in case of violation of the transmission condition.

The evaluation of the transmission energy and transmission function is much simpler if the analyzing point remains at 
${\bf P_A^0}$ for all starting angles. In that case, if we assume that the field lines are symmetric to
the $z=0$ mirror plane, 
all analyzing points are at the $z=0$ mirror plane (this is then also called analyzing plane). In addition,
the transmission function is determined by the electric potential and magnetic field  in the source and the analyzing plane only.
To satisfy the ${\bf P_A}={\bf P_A^0}$ transmission condition (independence of ${\bf P_A}$ from the starting angle), 
we have two possibilities in the layout of the electromagnetic fields. 
First, we can improve the homogeneity of the magnetic field
near the mirror plane, so that the change of
the third magnetic term in eq. \ref{Elong} becomes smaller than the
change of the second electric term. One could also make the electric potential near the mirror plane more inhomogeneous,
but then the potential will also be more inhomogeneous in  radial direction of  the analyzing plane, and this would be disadvantageous
for the precise determination of the transmission function.
Second, we can use a coil configuration where the magnetic field in the mirror plane
does not have a global minimum, but a local maximum instead \cite{aSPECT}
and  two local minima somewhere near the mirror plane.
In that case, when moving away from the point ${\bf P_A^0}$, 
the third magnetic term in eq. \ref{Elong} first  increases due to the decreasing
magnetic field, so that the longitudinal energy also increases. Farther away from ${\bf P_A^0}$ the magnetic field term decreases, but
there the second inhomogeneous electric potential term is able to overcompensate the magnetic term.
Accordingly, the analyzing point remains in the mirror plane
for all starting angles and magnetic field lines. 
We say that in these two cases the transmission condition is fulfilled.

\section{Physical requirements on the magnetic field in the main spectrometer}
 \label{SecRequirements}

To optimize the background and transmission properties for the KATRIN experiment, the magnetic field in the main spectrometer has to
fulfill certain requirements.

\begin{itemize}
 \item 
{\bf Magnetic guidance}

A key task of the magnetic field is to guide the electrons from the source to the detector
without electron trajectories  touching  beam line elements, as this
would result in a  loss of neutrino mass measurement statistics and in increased background.
For this reason, it is required that the flux tube (the 
bundle of  magnetic field lines  originating from the source) 
should fit well inside the main spectrometer tank.
Fig. \ref{FigFluxtube} shows the magnetic field lines corresponding to the boundaries of the 
reference 191 ${\rm Tcm^2}$ flux tube, without the LFCS air coils at the main spectrometer, with only the
stray fields of the s.c. solenoids (see Table \ref{TabSuper}).
The left figure a, includes the influence of 
the earth magnetic field, while the right figure b, assumes that the earth field is fully compensated.
In case a, the flux tube is strongly deformed by the earth field, so that a large part of the $\beta$-decay electrons
from the source would hit the inner walls of the spectrometer and thus  would not be detected.
In addition, secondary electrons from cosmic muon interactions would be guided to the detector, thus
increasing the background over  a large part of the flux tube.
 In case b, the earth magnetic field is 
assumed to be fully compensated, but 
the flux tube is still larger than the spectrometer, causing similar problems. 
Clearly,  an additional field shaping element is required to constrain the maximal diameter of the flux tube
so that it fits into the vessel geometry.

\item
{\bf Transmission condition}

The magnetic field configuration generated by the superconducting coils is asymmetric (has no mirror symmetry),
so that the transmission condition discussed in Sec. \ref{SecAdiabtrans} is not fulfilled. This calls for a field-shaping element which
allows to compensate the violation of the transmission condition.

\item
{\bf Homogeneity}

Once the transmission condition is satisfied, the magnetic field values within the analyzing plane 
should be as homogeneous as possible, so that  the  transmission function 
can be determined very precisely.

\begin{figure}[htbp]
\centering
\subfigure[without air coils]{\includegraphics[width=0.47\textwidth]{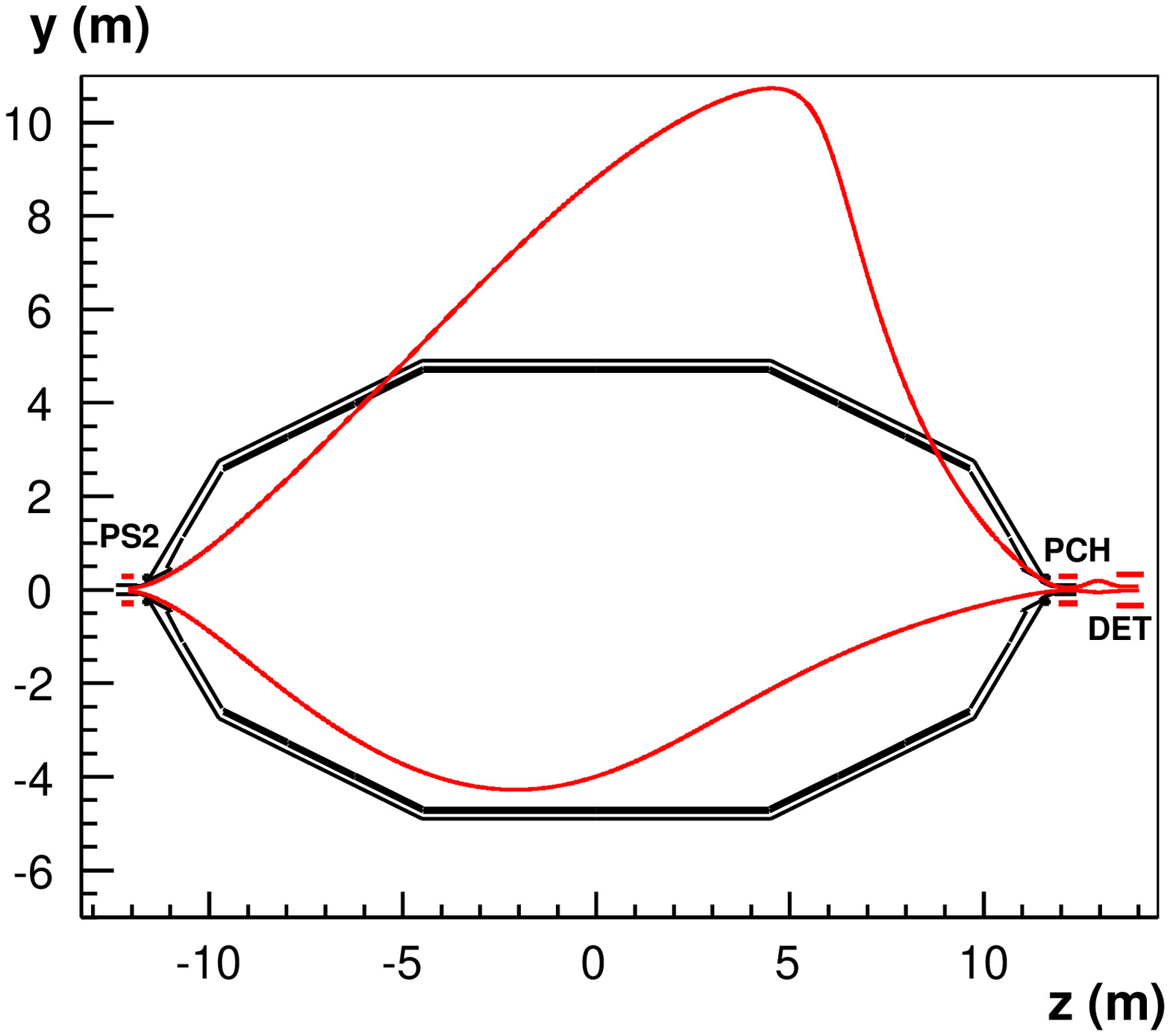}\label{FigFluxtube1}}\quad
\subfigure[with fully compensated earth magnetic field]{\includegraphics[width=0.47\textwidth]{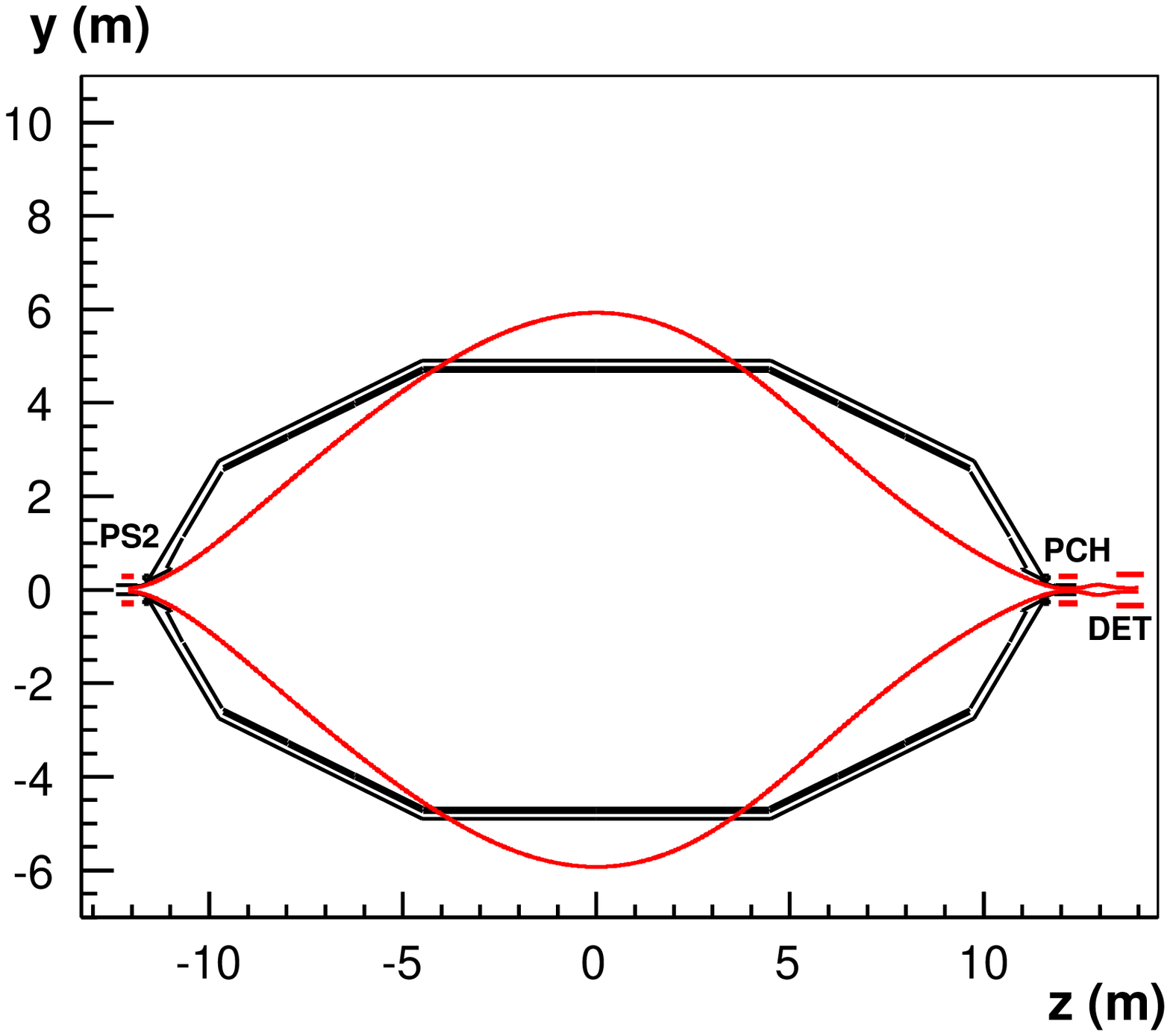}\label{FigFluxtube2}}         
\caption{The reference 191 ${\rm Tcm^2}$ flux tube magnetic field lines inside the main spectrometer 
without LFCS air coils.
Displayed is the precise vessel geometry with a steep and flat cone ends and a cylindrical middle part.
In case a, the influence of the earth magnetic field has not been compensated, while in case b, it has been 
assumed to be fully compensated.
The $y$ axis corresponds to the vertical direction.
For visualization purposes, the radial thicknesses of the coils in these figures are larger than the real values.
}
\label{FigFluxtube}
\end{figure}

\item
{\bf Background}

The magnetic field inside the main spectrometer is of key importance to minimize the cosmic ray $\mu$-induced background.
Previous investigations performed with
the Mainz neutrino mass spectrometer  \cite{Goldmann,Muller} 
and the KATRIN pre-spectrometer  \cite{Lammers,Groh} have revealed that  the background is smaller 
when the magnetic field inside
the spectrometer is higher. 
Namely, secondary electrons emitted at the inner surface of the spectrometer and electrodes  cannot easily
 move perpendicularly to magnetic field lines (they move much easier parallel to these field lines). 
Accordingly,  the magnetic field acts as strong shielding
against these electrons. For higher values of the magnetic field inside the spectrometer volume the shielding is more
 efficient, as for example the flux tube then is 
farther away from the inner tank and electrode surface. 
It is then also easier to fulfil the transmission condition, as
the electric potential is usually more inhomogeneous closer to the spectrometer axis.
On the other hand, a higher magnetic field at the analyzing plane reduces the energy resolution, thereby making 
the transmission function broader. As a result,
 we get a somewhat smaller signal rate and we have to know more precisely the transmission function.
Obviously, one has to find some optimum magnetic field with small background rate and acceptable energy resolution.
In addition, a good compensation of the earth magnetic field makes the overall magnetic field in the spectrometer more axially symmetric,
and this could also be important  to reduce the background. In addition, electron tracking simulations
indicate that the background could depend also on the magnetic field shape in the main spectrometer
(e.g. one minimum or two minima with local maximum) \cite{Gluck2005b} .

\end{itemize}

Taking into account the above considerations, an additional field-shaping element is required to guarantee
an optimized performance of the main spectrometer. This element is the  large volume air coil system
 surrounding the main spectrometer. The system combines two distinct units: the Low Field
Correction System (LFCS) to fine-tune the axisymmetric low field part of the magnetic guiding flux tube, 
and the Earth Magnetic field Compensation System (EMCS). Both systems are described in detail below.

\section{The Low Field Correction System (LFCS)}
 \label{SecLFCS}

\vspace{2mm}
{\bf General overview}
\vspace{2mm}

\begin{figure}[htbp]
\centering
\includegraphics[width=0.90\textwidth]{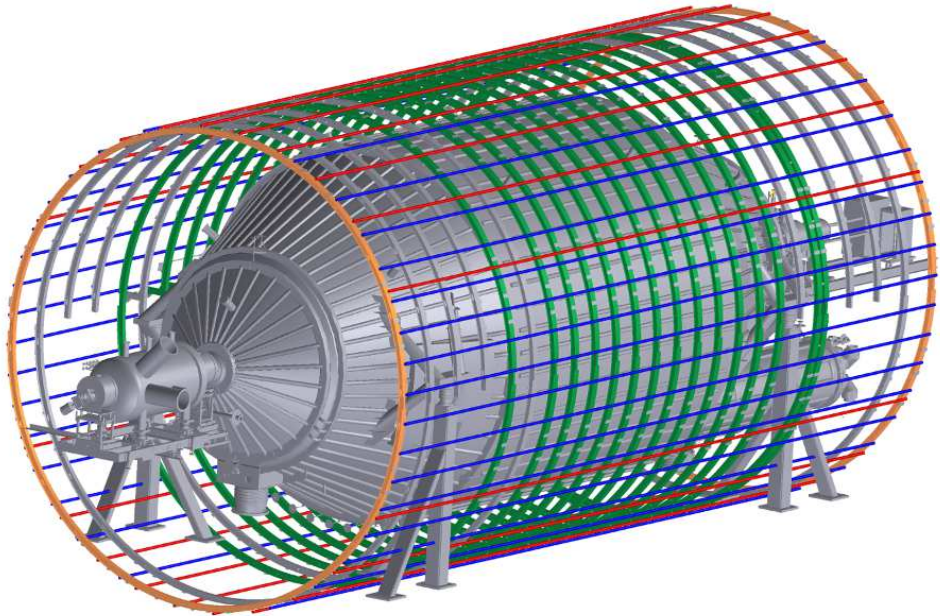}
\caption{The KATRIN main spectrometer with the large volume air coil system.
The green circles represent the LFCS coils. The blue and the red straight lines
belong to the current loops of the vertical and the horizontal components of the EMCS, respectively. 
The two orange circles
at the ends contain the current arcs that connect together the linear current sections of the EMCS.
}
\label{FigMainspec}
\end{figure}

The LFCS   comprises 14 large (12.6 m diameter) air coils arranged
coaxially with the main spectrometer tank and  the superconducting coils at both ends of  the spectrometer
(the green circles in Fig. \ref{FigMainspec}).
Table \ref{TabLFCS}  lists the axial coordinates $z_c$, winding numbers $N_{\rm turns}$
and maximal currents $I_{\rm max}$
of the LFCS coils.
Due to  the large gaps between the neighbouring coils
(70 cm or more), there is enough space for accessing various parts of the
main spectrometer tank from outside the air coil system.
Each coil is driven by its own power supply, so that the currents in each coil can be adjusted
individually. As a result,   different magnetic field profiles
inside the main spectrometer can be implemented on short time scales.
This allows for precision fine-tuning of the shape of the magnetic
field and for adjusting the total magnetic field strength to  various
needs.

\begin{table}
\begin{center}
\vspace{2mm}
\begin{tabular}{| l  | c   |  c  |  c   |  c  | c |} \hline
Coil index   &  $z_c$ (m) &   $N_{\rm turns}$ &  $I_{\rm max}$ (A)   &$I$ (A) [1 min.] &  $I$ (A) [2 min.]  \\ 
\hline \hline
1                 &      -6.79     &         14         &   100           &     -11.2                    &       -0.5              \\
\hline
2                 &     -4.94      &         14         &    100          &       -15.3               &        0.            \\
\hline
3                 &      -4.04     &          8         &      125         &        -7.9             &        -4.8           \\
\hline
4                 &      -3.14      &        8         &       125          &       -13.4              &        -7.1           \\
\hline
5                 &      -2.24       &       8         &      125           &        -12.2             &       -6.6           \\ 
\hline
6                 &      -1.34       &        8       &       125           &        -24.2             &       -19.4          \\ 
\hline
7                 &        -0.44      &       8       &       125             &        -17.1              &       -57.2           \\ 
\hline
8                 &         0.46      &       8        &      125            &         -20.3              &       -51.2           \\ 
\hline
9                 &         1.35      &       8       &      125             &         -18.5              &      -22.7            \\ 
\hline
10               &         2.26       &     8        &      125            &          -23.1             &       -12.5           \\ 
\hline
11               &         3.16        &    8        &       125           &         -21.9             &       -7.7           \\ 
\hline
12               &         4.06         &   14       &      100          &         -18.1              &       -16.8           \\ 
\hline
13               &         4.95         &    14       &     100          &         -13.3              &       -15.9           \\ 
\hline
14               &  6.6 and   6.9   &   14 + 14    &  70      &          27.3             &       42.1           \\ 
\hline
\end{tabular}
\caption{Optimized LFCS coil currents for 2 different field configurations: 1 global minimum (column 4),
and 2 local minima  with a local maximum for off-axis field lines (column 5).
$z_c$ is the axial position of the coil center, $N_{\rm turns}$ denotes the number of  turns,
and $I_{\rm max}$ is the maximum current of the constructed coil.
Coil 14 is implemented as a double coil, therefore it has 2 different axial positions; both subcoils have 14 turns.
All LFCS coils have 6.3 m inner radius, 2 cm radial thickness and 19 cm axial length.
\label{TabLFCS}} 
\end{center}
\end{table}

With the help of the LFCS coils
we can set the magnetic field value in the middle of the spectrometer to any value up to 1 mT. In normal neutrino mass measurement conditions
the magnetic field at the center ($z=0, r=0$) of the spectrometer 
should be larger than a minimum value of
0.33 mT, so that the reference 191 ${\rm Tcm^2}$ flux tube fits 
into the spectrometer tank. The field of the superconducting coils at the center contributes with about 0.2 mT, therefore  the
overall magnetic field direction of the  LFCS has to be the identical to the field direction of the superconducting coils. 
If a higher magnetic field in the middle of the main spectrometer is required,  the background
level is expected to be reduced significantly, but then the transmission function is
broader, i.e. the energy resolution is worse.

\begin{figure}[htbp]
\centering
\subfigure[1 minimum field configuration]{\includegraphics[width=0.70\textwidth]{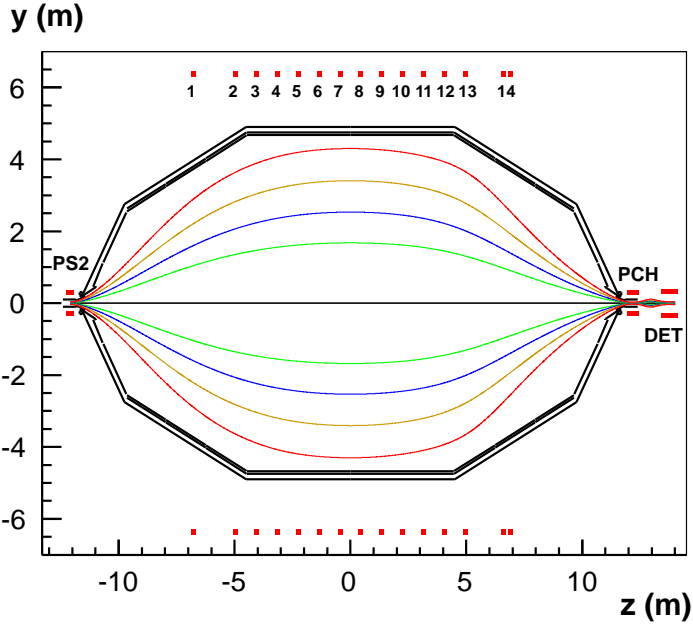}\label{FigFieldlines1}}\quad
\subfigure[2 minima field configuration]{\includegraphics[width=0.70\textwidth]{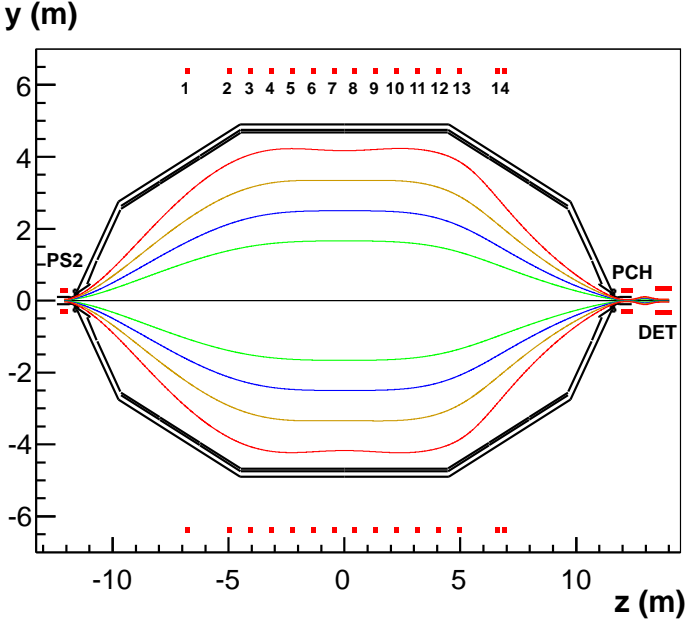}\label{FigFieldlines2}}         
\caption{
Magnetic field lines inside the main spectrometer (side view), with 0.35 mT field at center. 
The field lines correspond to the following magnetic flux values (with increasing distance from the axis):
0 ${\rm Tcm}^2$ (black), 30 ${\rm Tcm}^2$ (green), 68 ${\rm Tcm}^2$ (blue), 122 ${\rm Tcm}^2$ (yellow),
191 ${\rm Tcm}^2$ (red).
For visualization purposes, the radial thicknesses of the coils in these figures are larger than the real values.
}
\label{FigFieldlines}
\end{figure}

As outlined above, the field of the superconducting coils is rather asymmetric
with regard to the middle plane, since their stray field at the 
detector side of the main spectrometer is larger than at the source side (see also Fig. \ref{FigFluxtube2}).
With the LFCS it is possible
to compensate  this asymmetry to a large extent.
 For this purpose, the LFCS coil 14 
at the detector side (see Fig. \ref{FigFieldlines}) will  be used as a counter coil  with
a current direction opposite to all other coils. 
As this task requires a rather large amperturn value, 
coil 14 consists of 2 parts, each of them having 14 windings with slightly different axial coordinates
(see Table \ref{TabLFCS}).
 
The currents of the LFCS coils have to be optimized so that the transmission condition is fulfilled.
This task can best be realized if the superconducting stray field is smaller  and the LFCS field is larger.
Accordingly, we define the 
KATRIN magnetic field direction opposite to the horizontal earth magnetic field direction.
In this layout, the earth field reduces the stray field of the superconducting coils, as desired.
As outlined earlier, a big advantage in this regard is that
the main spectrometer axis has approximately a
south-to-north direction (detector side is at north).
Accordingly, the axial ($z$) component of the earth magnetic field has source-to-detector
direction (20 $\mu$T; see table \ref{TabSuper}). This allows  
to choose  a detector-to-source (negative) direction for the KATRIN magnetic field, in order
to reduce the superconducting stray field by the horizontal earth field.

\vspace{2mm}
{\bf Field optimization}
\vspace{2mm}

As we have mentioned in the beginning of this section, the LFCS coils allow to
adjust many different magnetic field configurations: with various field magnitude values up to 1 mT,
with one minimum or two minima field solutions, and with different superconducting fields.
In our paper, we present two generic field configurations: first, a configuration with
one global magnetic field minimum for all field lines,
and second, a configuration with 
 2 local minima and a local maximum for off-axis field lines. In both cases, the field at the center of the
main spectrometer is 0.35 mT. For these calculations we have included the contributions from all superconducting
coils of the KATRIN system. Table \ref{TabSuper} shows the central axial positions,  central fields and field contributions
at the main spectrometer center for the 7 superconducting coil systems and coils.

\begin{figure}[htbp]
\centering
\subfigure[1 minimum field configuration]{\includegraphics[width=0.47\textwidth]{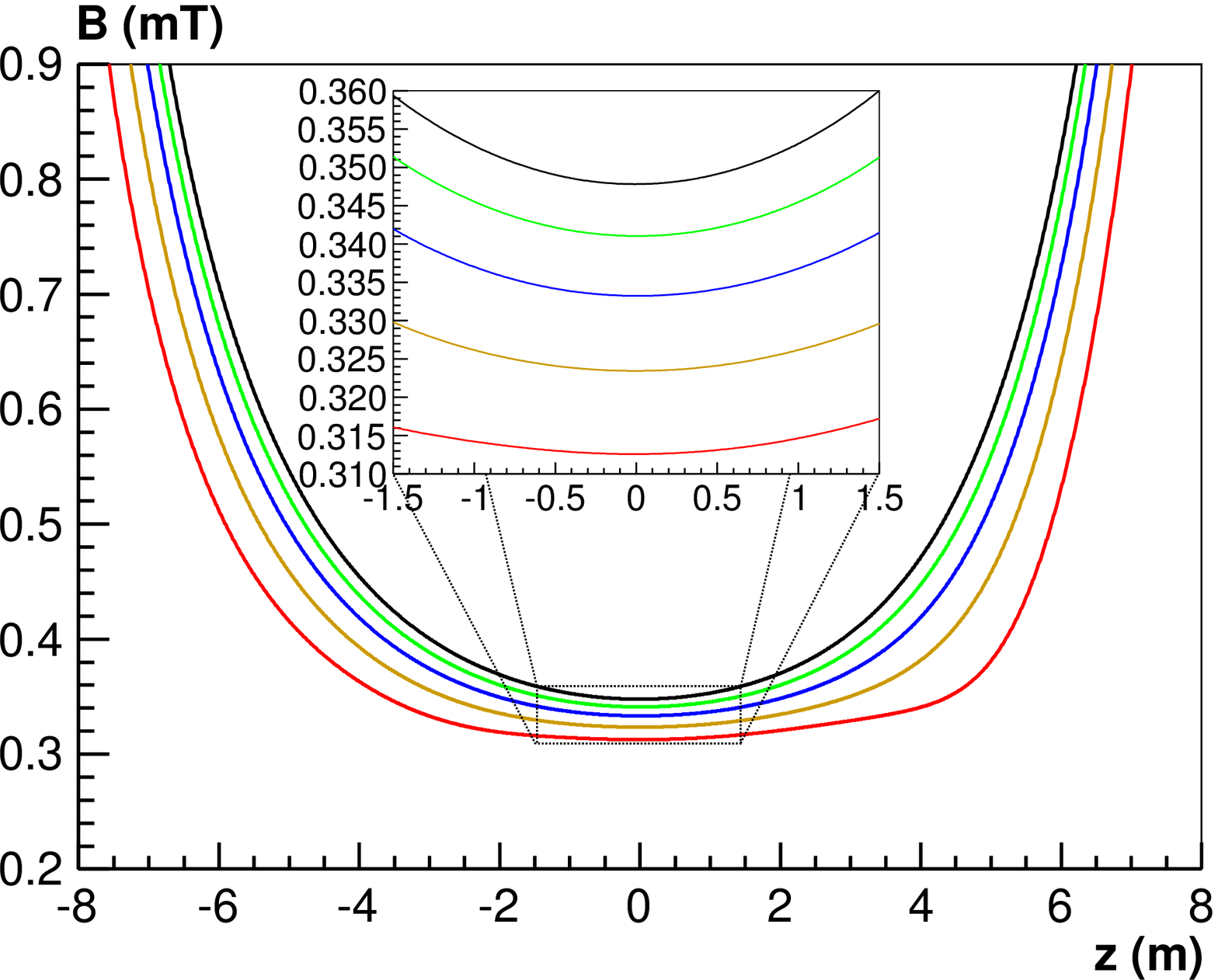}\label{FigMagfield1}}\quad
\subfigure[2 minima field configuration]{\includegraphics[width=0.47\textwidth]{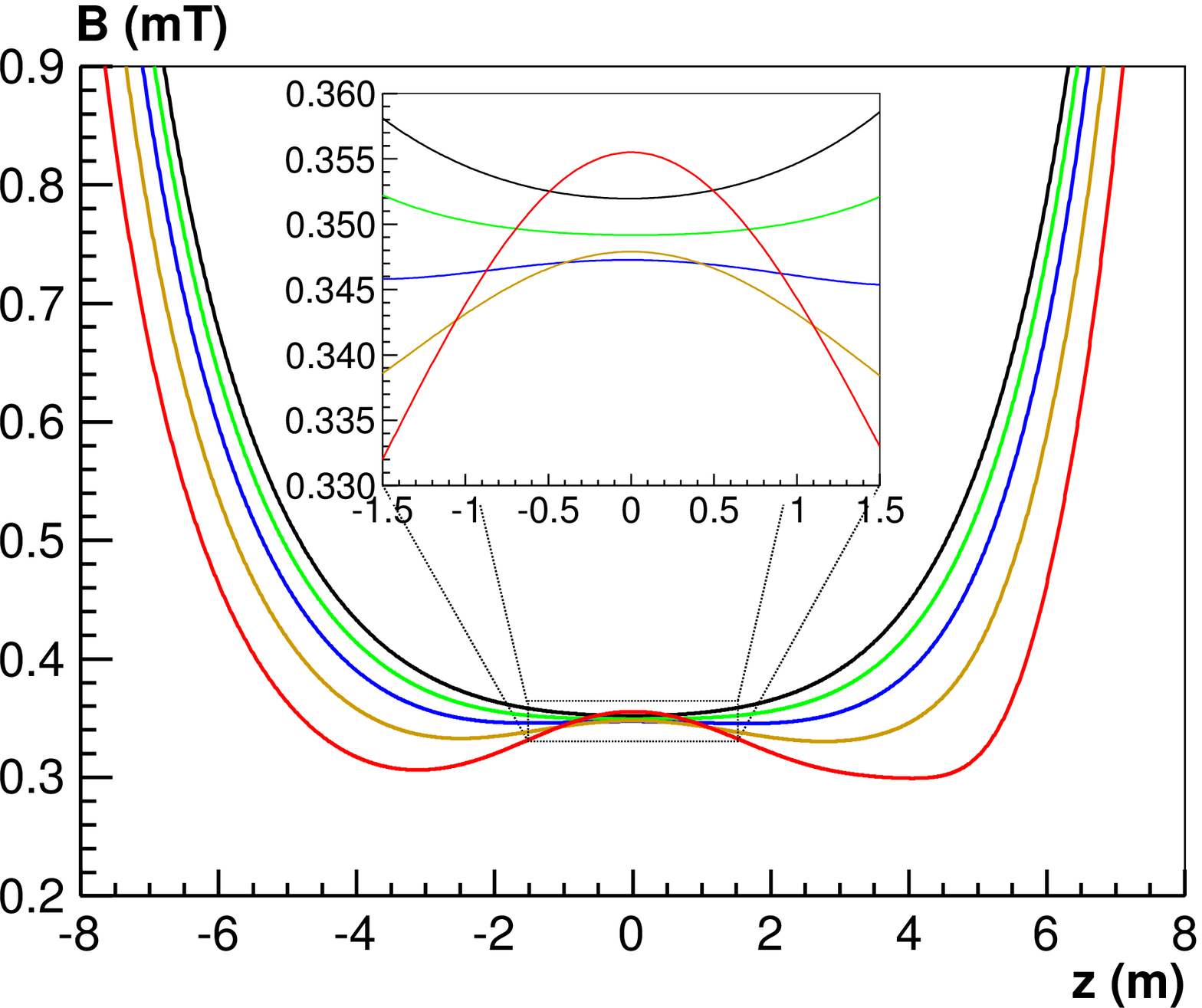}\label{FigMagfield2}}         
\caption{Magnetic field along the field lines of Fig. \ref{FigFieldlines}, with 0.35 mT field at center. 
The colors here correspond to
the field line colors of Fig. \ref{FigFieldlines}.
The inlets show the field strength in the innermost part in more detail.}
\label{FigMagfield}
\end{figure}

\begin{figure}[htbp]
\centering
\subfigure[1 minimum field configuration]{\includegraphics[width=0.47\textwidth]{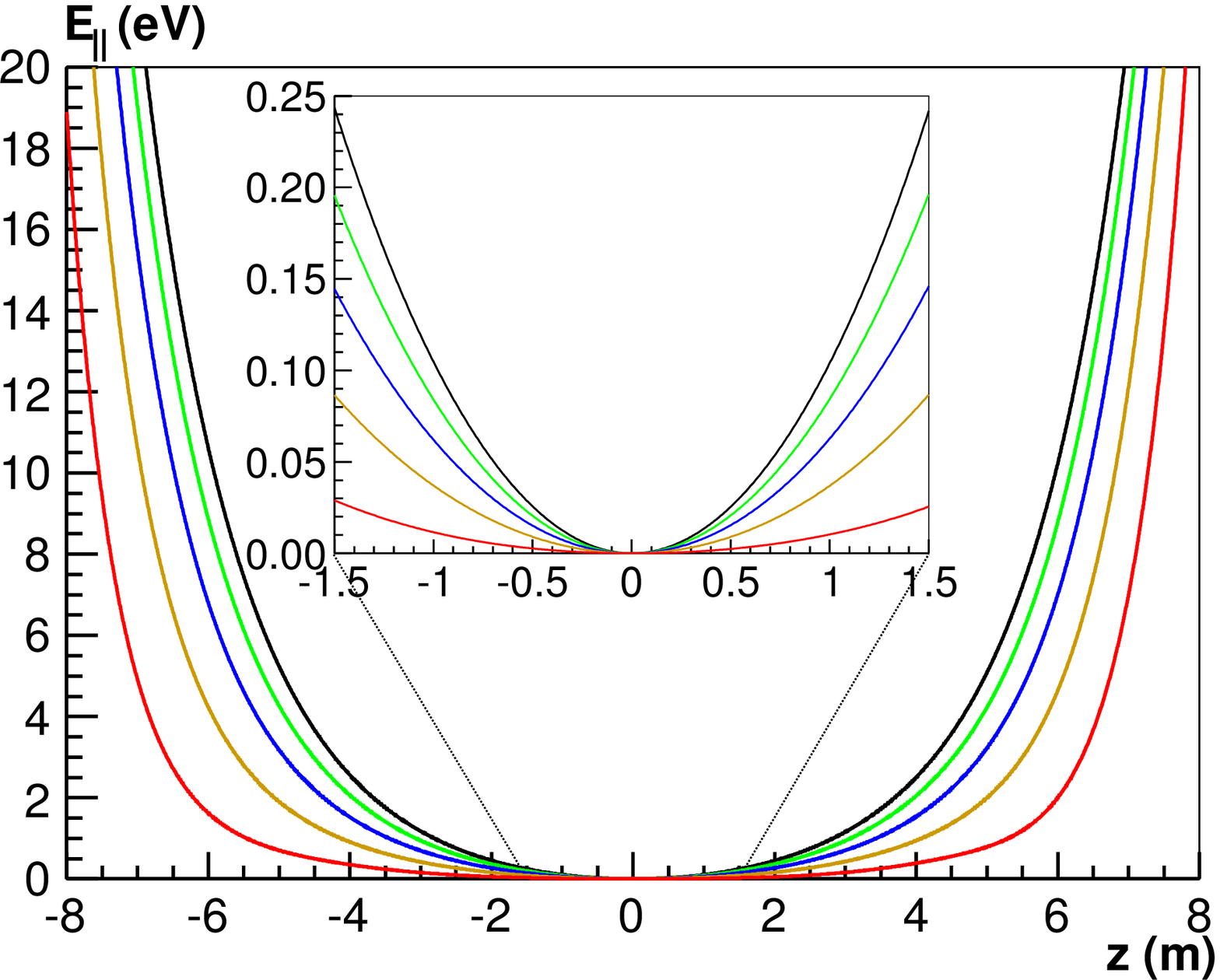}\label{FigElong1}}\quad
\subfigure[2 minima field configuration]{\includegraphics[width=0.47\textwidth]{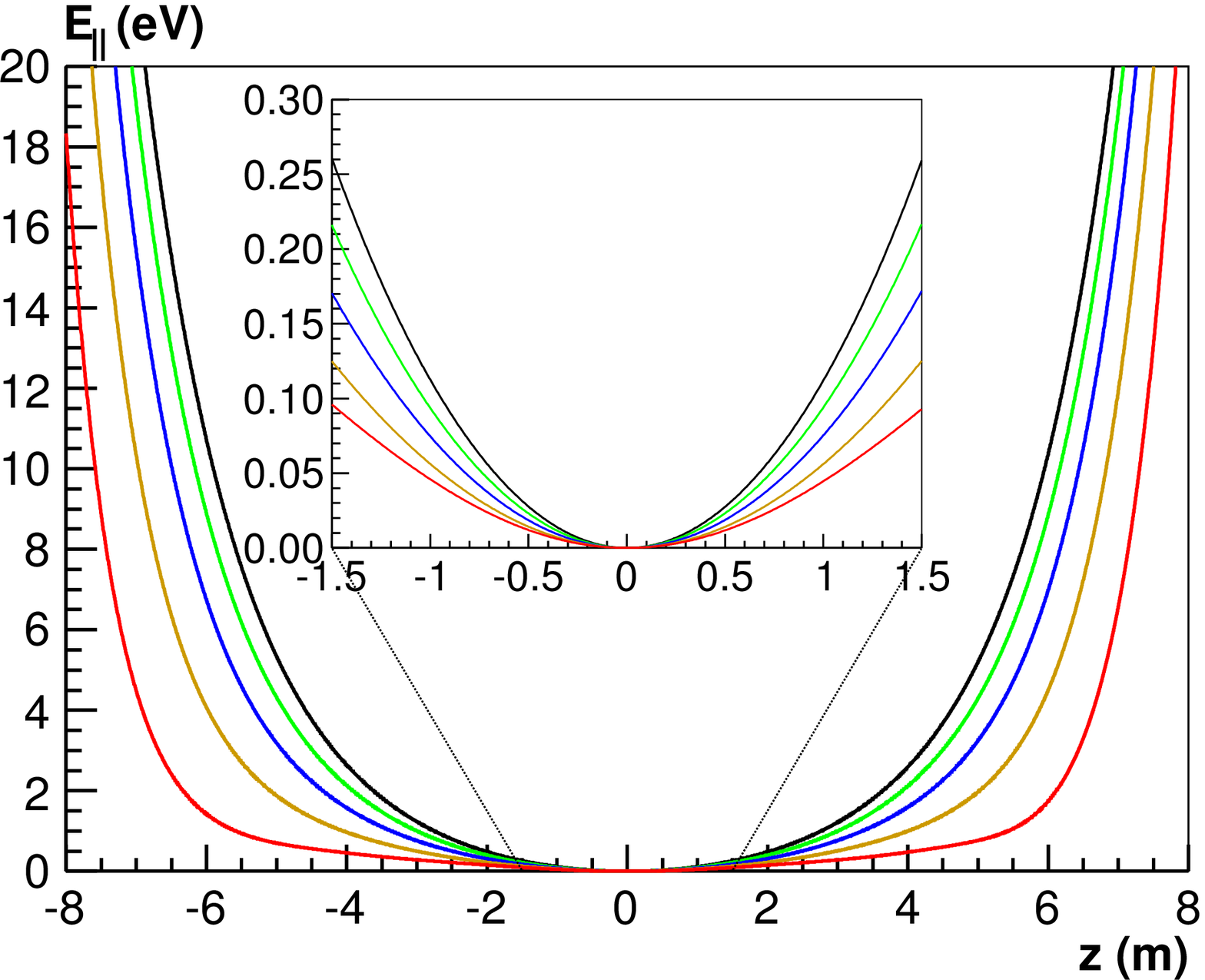}\label{FigElong2}}         
\caption{Longitudinal energy distribution along the magnetic field lines of Fig. \ref{FigFieldlines},
 with 0.35 mT field at center. 
The colors here correspond to
the field line colors of Fig. \ref{FigFieldlines}. The electrons were started in the source (WGTS) with transmission energy and with
 maximal polar angle $51^\circ$.
For these simulations, most of the wire modules were on vessel potential, except of the two smallest wire module rings 
at the steep cone part of the main spectrometer, which
were 200 V and 300 V more postive than the vessel.
}
\label{FigElong}
\end{figure}

In order to determine the optimized LFCS coil currents, we have used an optimization procedure
based on  reasonable initial values for the currents. Then we computed the magnetic field lines inside
the flux tube (Fig. \ref{FigFieldlines}), as well as  the magnetic field  (Fig. \ref{FigMagfield}) 
and the adiabatic longitudinal energy (Fig. \ref{FigElong}) of an electron along these field lines.
Note that these figures correspond to the final optimized current values; in the initial stages of
our optimization simulations
the parameters  looked differently. For example, it occurred that
the outer field lines crossed the main spectrometer tank or electrodes, so that
one had to increase the absolute value of the LFCS coil currents to increase the magnetic field inside the main spectrometer.
If the field lines had a too large diameter in some local region, 
one had to increase the current values only for the coils near that region. 
In addition, an important design goal was to set the LFCS currents so that the magnetic field along the field lines
(Fig. \ref{FigMagfield})  is approximately symmetric relative to the $z=0$ plane: in that case one has better chances
to fulfill the transmission condition. The latter could be tested by the longitudinal energy figures (Fig. \ref{FigElong}).
After a few iterations of changing the current values, it was possible to find a configuration which
 approximately fulfilled the above criteria.

However, after this so called optimization-by-eye procedure, 
the analyzing points for various starting points and starting angles still had some spread
(lying within a region of a few times 10 cm size). In order to reduce substantially the distances between these analyzing
points, we used a mathematical optimization method  based on several objectives (multiobjective optimization),
minimizing the composite objective function by the downhill simplex method  \cite{NelderMead,Press}. 
The objective function depends on the 14 LFCS coil currents, therefore the optimization proceeds in this case
in a 14 dimensional parameter space.
The results of the optimization-by-eye method served as useful starting points for the mathematical
optimization procedure.
In this way we were able to improve significantly the transmission properties of the field configurations
(see below).
We give a detailed explanation of our
mathematical optimization method in \ref{appendix-mathoptim}.

\vspace{2mm}
{\bf Results for two field configurations}
\vspace{2mm}

\begin{figure}[htbp]
\centering
\includegraphics[width=0.60\textwidth]{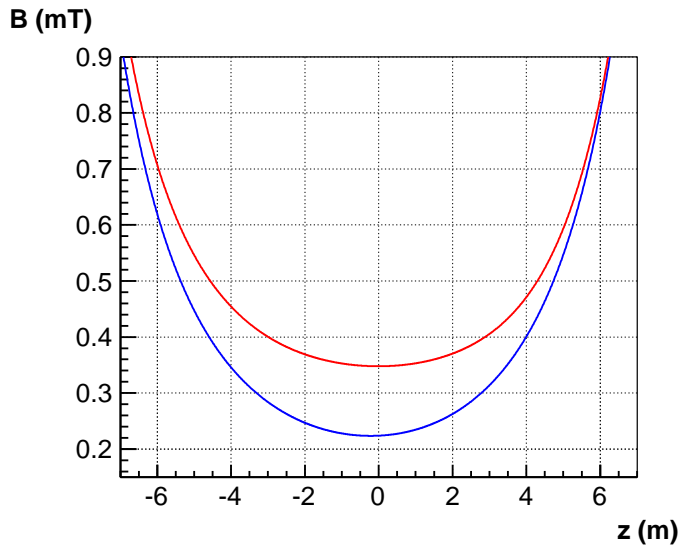}
\caption{Absolute value of the on-axis magnetic field. 
The position $z=0$ corresponds to the center of the main spectrometer.
Lower (blue) curve: field of superconducting coils and horizontal earth field
alone; upper (red) curve: same as for lower curve, but now the optimized LFCS coils (1 minimum case) have also been included.}
\label{FigMagfieldsuper}
\end{figure}

Table \ref{TabLFCS} shows the resulting LFCS coil current values for the two field configurations
(1 minimum and 2 minima), based on  the  abovementioned optimizations.
In both cases, the current of coil 14 is  positive, i.e. opposite to the sign of all other superconducting and LFCS air coils.
In this way the LFCS coil 14 can compensate (at least in the smaller field region) the 
asymmetry resulting from the larger stray fields of the 
pinch and detector coils.
Fig. \ref{FigMagfieldsuper} shows that the on-axis field without the LFCS coils is larger in the positive $z$
region (detector side); with the LFCS coils this asymmetry becomes smaller.
 It is also noticeable in table \ref{TabLFCS} that for the 2 minima configuration the central coils 7 and 8 have 
to be operated with rather large
current values, because
in this case the magnetic field is designed to have a local maximum at the center ($z=0$)
of the off-axis field lines.

Fig. \ref{FigFieldlines} shows the shape of 5  selected magnetic field lines for both field configurations
(1 minimum and 2 minima) after the abovementioned optimizations.
In each case the outer field lines  correspond to the  191 ${\rm Tcm}^2$ flux tube.
One can see that  the flux tube fits well into the spectrometer tank, 
in either case with some safety distance (about 40 cm) from the 
inner wire electrode. Close to the analyzing plane the field lines
 display a high degree of symmetry relative to the $z=0$ plane, while
farther away at the detector side the field 
lines attain a smaller diameter than at the source side (due to the higher stray field of the pinch and detector coils).
The LFCS coils are also indicated in these figures by the points at $y=6.3$ m and $y=-6.3$ m. 
Only those 3 superconducting coils are displayed here which are closest to the main spectrometer
(the PS2, PCH and DET coils at $z=-12$ m, $z=12$ m and $z=13.8$ m, respectively).

Fig. \ref{FigMagfield} displays  (in an identical color code)
the magnetic field  strength along the 5 selected magnetic field lines of Fig. \ref{FigFieldlines}.
Each field line obeys a very good approximate symmetry relative to $z=0$ 
(although the field of the superconducting coils alone is quite
asymmetric in $z$-direction), implying that the compensation by the LFCS  is  successful. 
In case of the 1-minimum configuration the
field has only 1 rather shallow minimum at $z=0$, while 
for the ``2-minima`` layout this only manifests
for the inner field lines (upper curves); 
 the outer field lines (lower curves) 
experience 2 local minima (a few meters far from the center) and a local maximum at $z=0$.
As  explained before, the latter configuration is more reliable from the transmission condition point-of-view.
A possible disadvantage of these local field minima is that some electrons with velocities almost perpendicular 
to the magnetic field could be trapped in these minima. In this way they could cause background by
ionizational collisions, in case of sufficient kinetic energy. 
However, these electrons will be stored anyway by the magnetic mirror trap of the main 
spectrometer magnetic field, therefore it is unlikely that
the local field minima would result in a significant background increase, in comparison to the expected background rate due to
the ionizations caused by high energy stored electrons in the main spectrometer \cite{Mertens2013}.
Of course, this specific background issue has to be investigated  experimentally.

\begin{figure}[htbp]
\centering
\subfigure[1 minimum field configuration]{\includegraphics[width=0.47\textwidth]{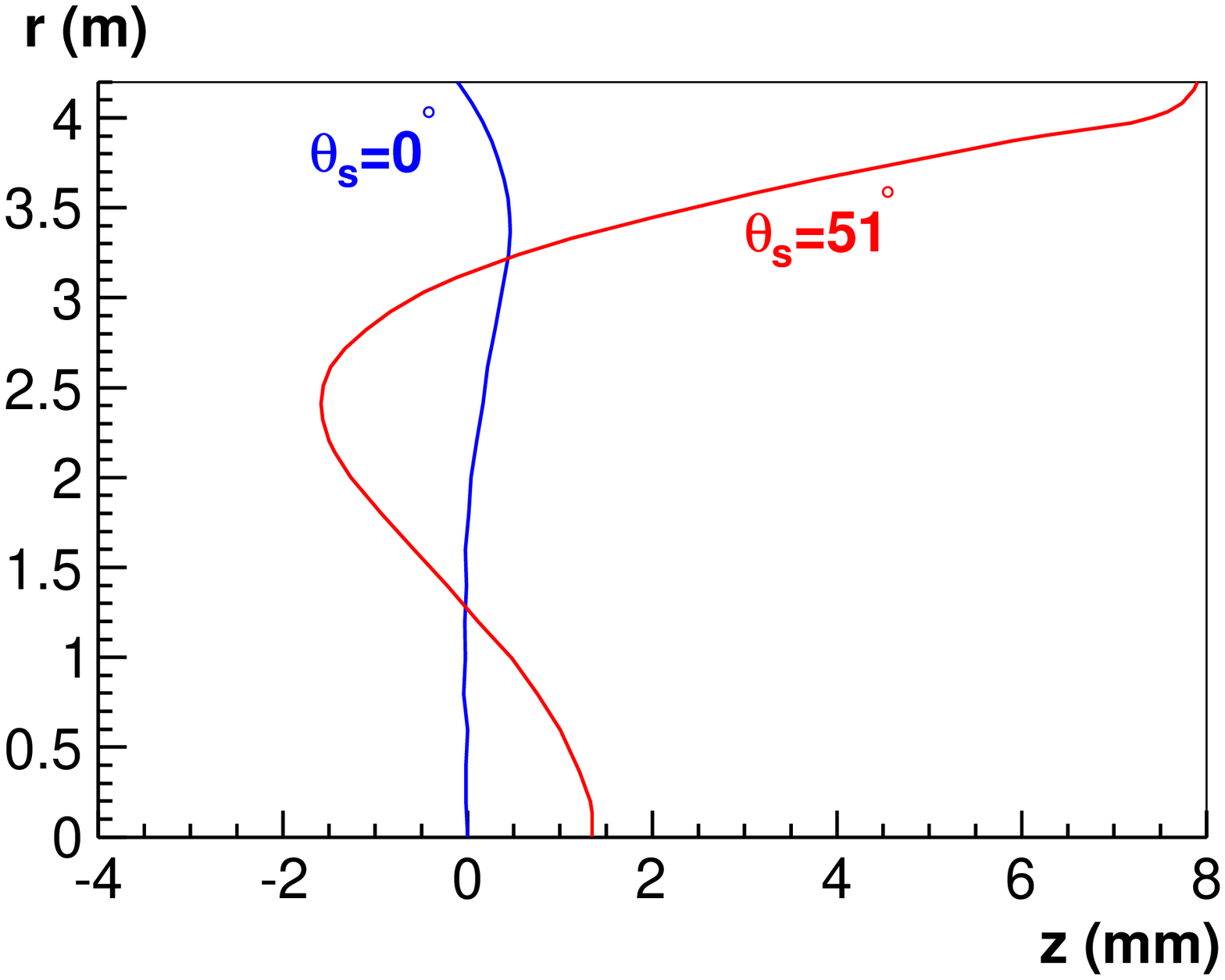}\label{FigAnapoint1}}\quad
\subfigure[2 minima field configuration]{\includegraphics[width=0.47\textwidth]{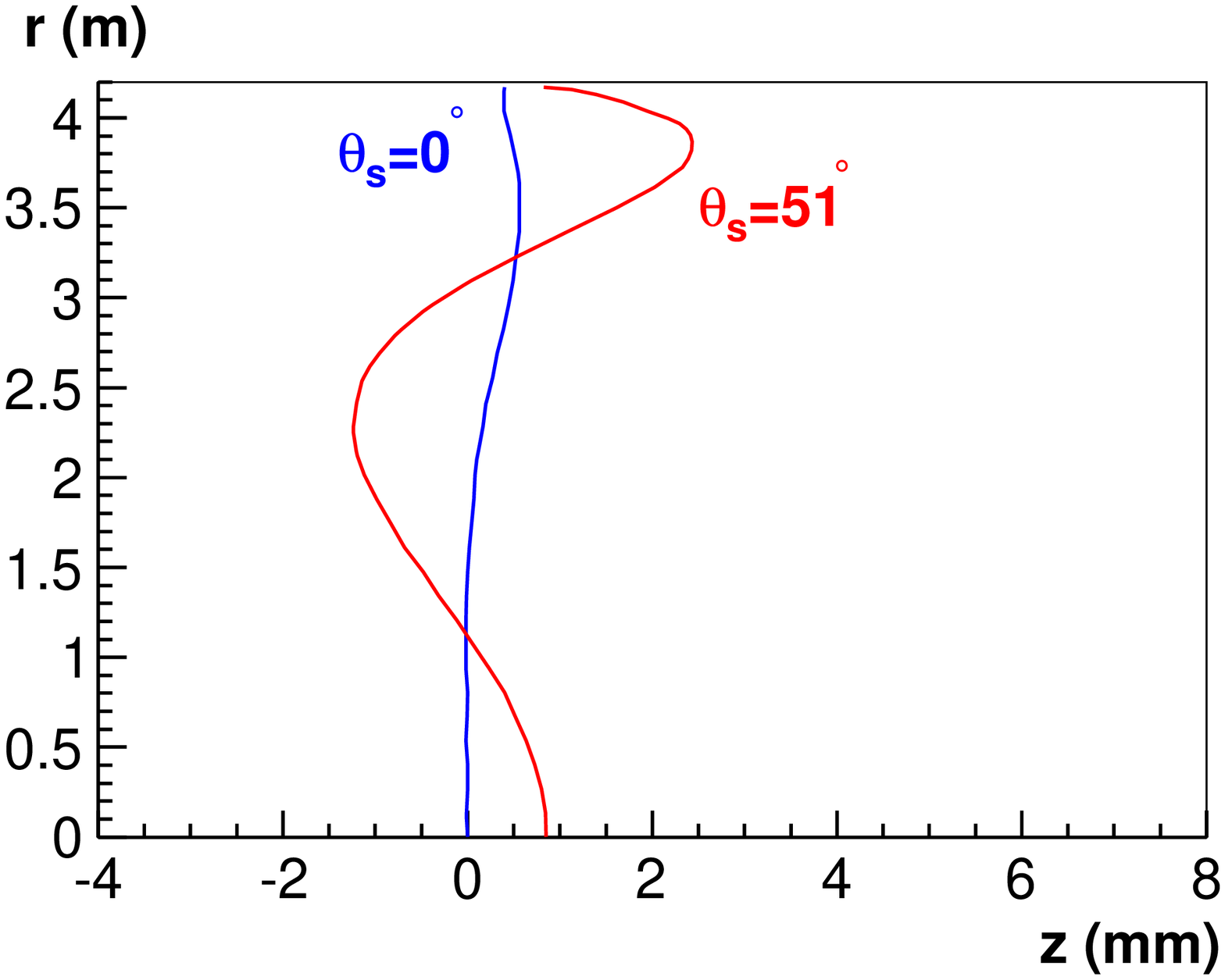}\label{FigAnapoint2}}         
\caption{The $z$ and $r$ coordinates of the analyzing points for various field lines and for
the minimal and maximal  starting polar angles ($\theta_s=0^\circ$ and $\theta_s=51^\circ$).
Note the scale of the $z$-axis in mm.
}
\label{FigAnapoint}
\end{figure}

Fig. \ref{FigElong} shows the longitudinal energy along the same field lines, for electrons starting with the maximal
polar angle  $51^\circ$  and with the transmission energy 
$E_s=E_{\rm tr}$ in the source (WGTS).
In both cases, the minimum of the longitudinal energy appears very close to the $z=0$ symmetry point.
To zoom into the critical region-of-interest, we display
 in Fig. \ref{FigAnapoint}  the analyzing points for the minimal and maximal starting polar angles
($\theta_s=0^\circ$ for the blue lines, $\theta_s=51^\circ$ for the red lines),
 but now with a mm scale on the $z$-axis,
for all field lines with $r<4.2$ m at the center.
The analyzing points for intermediate starting angles  lie between these two curves.
This figure demonstrates that
our optimization method for both field configurations results in a very small axial spread of the
analyzing points, typically on a scale of a few mm only.
 It is  not meaningful to further improve   these analyzing point curves,
 because  small magnetic and electric field disturbances  would change these results.
For example, the magnetic field of the coils is slightly disturbed by the presence of 
magnetic materials in the main spectrometer building,
and the mirror symmetry of the electric field is affected  by the detector-facing pumping ports
(which are  at the detector side of the main spectrometer), resulting in systematic effects of the same order of a few mm.  
Taken together, these results imply that the two generic field configurations described above result in
a well-defined analyzing plane with a narrow spread in the few mm range, as desired for high-resolution
$\beta$-spectroscopy.

\begin{figure}[htbp]
\centering
\subfigure[1 minimum field configuration]{\includegraphics[width=0.47\textwidth]{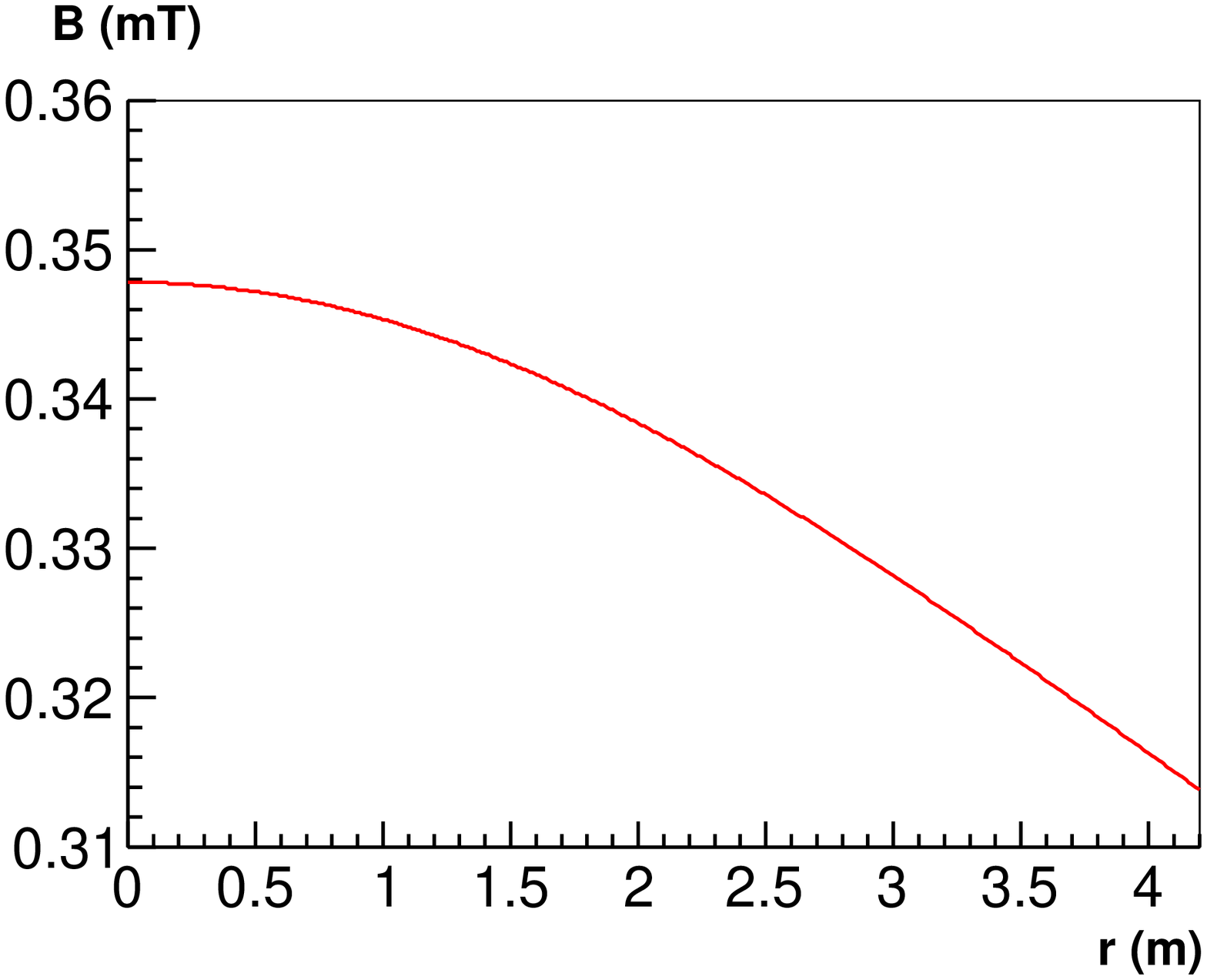}\label{FigMagap1}}\quad
\subfigure[2 minima field configuration]{\includegraphics[width=0.47\textwidth]{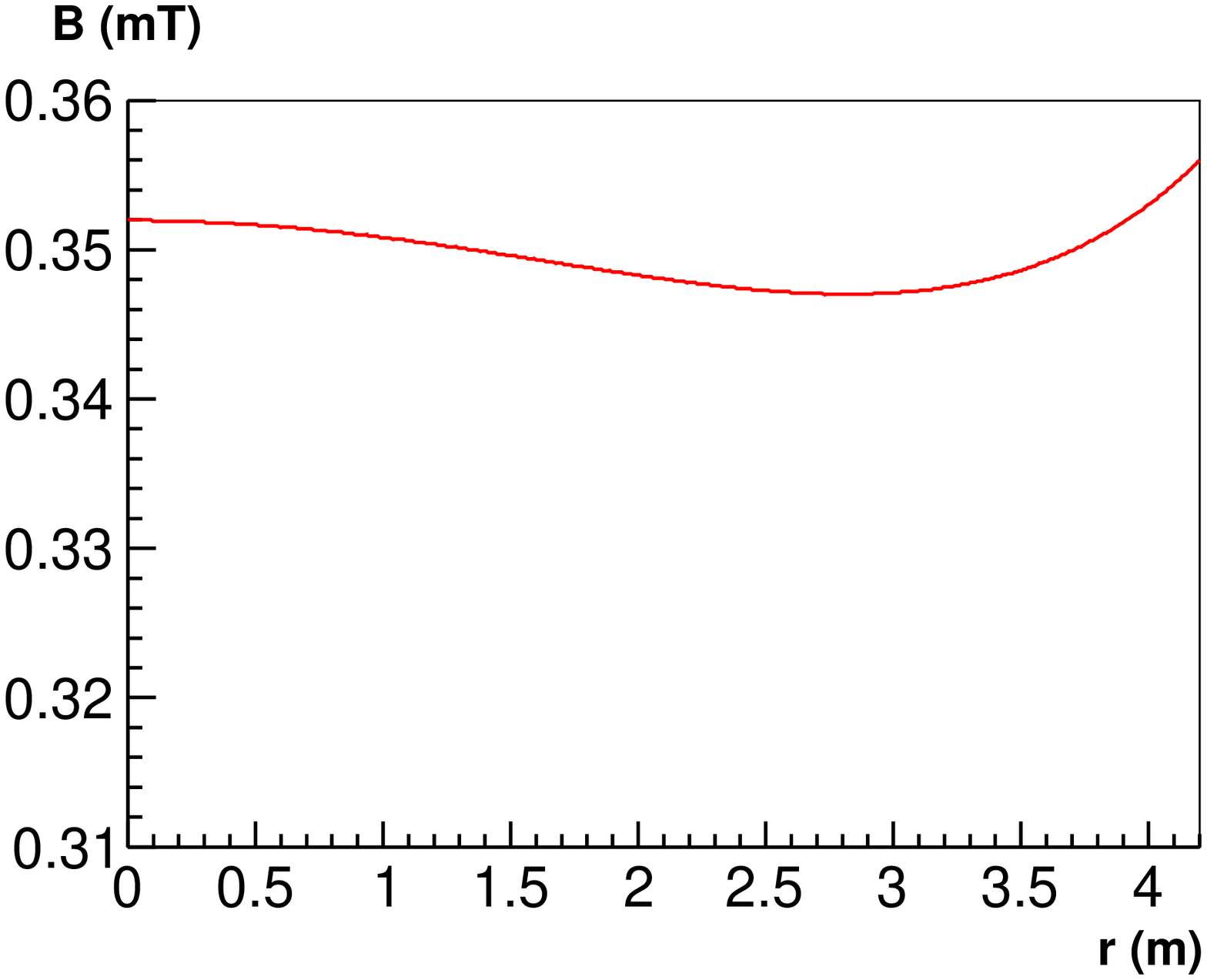}\label{FigMagap2}}         
\caption{Magnetic field in the analyzing plane ($z=0$), as function of the radius.
}
\label{FigMagap}
\end{figure}

The final important parameter to be investigated is
the radial homogeneity of the 
magnetic field in the analyzing plane ($z=0$).
In  Fig.  \ref{FigMagap} we can see clear differences of
the two field configurations: the LFCS setup with the 2 minima and local maximum offers a  significantly better radial
homogeneity than the field configuration with only 1 global minimum.
In principle, as outlined above, the magnetic inhomogeneity influences the energy resolution, so a better homogeneity
is advantageous. However, this effect can be mapped out to some extent
by the segmented focal plane detector \cite{Harms}.

In addition to these two examples of LFCS current setting, we have calculated several other current configurations: 
scenarios with a higher overall magnetic field in the analyzing plane, and starting configurations with
 2 or 4 superconducting coils only. The main spectrometer
test experiments, which will be performed in 2013, will use a configuration with
only the  PS1, PS2, PCH and DET
superconducting coils (the WGTS, DPS and CPS are at present still under construction).
For these reasons it is important to find optimal LFCS current configurations
with stray  magnetic fields caused by these  4 superconducting solenoids. One can find optimized LFCS current
values for these cases (with many figures) in Refs. \cite{Gluck2006,Gluck2009,Wandkowsky,Mertens2012,Gluck2012b}.

At the end of this section, we mention a possible application of the LFCS  for the purpose of background reduction.
As outlined in Sec. \ref{SecKatrin}, the MAC-E filter principle of
the main spectrometer inherently forms
 a magnetic mirror trap for high-energy electrons. These trapped electrons 
undergo many ionizational collisions with residual gas molecules, and the secondary electrons created by these ionizations can cause a significant
background increase \cite{Mertens2013,Mertens2012}. 
A possible method to remove these trapped electrons is by reducing the magnetic field in the middle of the main
spectrometer for a short time (e.g. 1 s) down to zero. 
This is possible by reversing the sign of the LFCS  currents. Using this 'magnetic pulse' 
method \cite{Otten,FurseMP},  
all high energy stored electrons are expected to be removed, and this should reduce significantly the background caused by these
electrons.

\section{The Earth Magnetic field Compensation System (EMCS)}
 \label{SecEMCS}

Since the earth magnetic field  is homogeneous within the volume of the KATRIN main spectrometer, 
it is possible to compensate this field distortion
 with the help of a homogeneous magnetic field. The widely known method to produce  such a homogeneous  field is
by circular or squared Helmholtz-type coil systems \cite{Fujita,Kirschvink,Sasada}, where the 
homogenous field region achieved
is, however,  significantly smaller than the dimension of the coil system itself. 
Since the building housing the KATRIN
main spectrometer and the LFCS system described above offers no extra space,
a Helmholtz-type coil method was not a viable method for the earth magnetic field compensation.

Another method to obtain a uniform magnetic field is by  spherical cosine coils \cite{Clark,Everett}. The layout of this system relies on
the fact that the magnetic field inside a uniformly magnetized sphere is uniform, with the induction vector ${\bf B}$
being parallel to the magnetization vector  ${\bf  M}$
\cite{Jackson}. 
From the point of view of the field intensity ${\bf H}$ calculations,
the uniform magnetization
${\bf  M}$ can be replaced by an equivalent surface current distribution
${\bf K}_m={\bf  M}\times {\bf  n}$,
where ${\bf  n}$ is the outwardly directed normal vector of the magnetic material
surface (see Refs. \cite{Cowan,PanofskyPhillips}).  Therefore, the equivalent current density is proportional to $\cos\theta_M$, where
$\theta_M$ is the angle between the normal vector ${\bf  n}$ and the plane perpendicular to ${\bf  M}$. This is the reason for
naming this arrangement a cosine coil. In order to build a cosine coil and 
to get an approximately uniform magnetic field  ${\bf  B}$ inside the coil,
the continuous current distribution is replaced
by a discrete system of circular current loops, with planes perpendicular to 
the axis vector ${\bf  B}$ and  positioned  equidistantly along the direction of this vector. 
As one can see in table 3 of Ref. \cite{Everett}, the spherical cosine coil system has a much larger
region with a specific level of field uniformity than the simple Helmholtz coil pair.

Unfortunately,  a spherical EMCS turned out to be impractical too, due to 
the above mentioned space restrictions
by the spectrometer building. Alternatively, a cosine coil system on the
surface of an ellipsoid also features a
uniform magnetic field inside the ellipsoid \cite{Clark,Smythe}. However, in this case the 3 
large vacuum pumps of the main spectrometer tank (see Fig. \ref{FigBeamline})
would have crossed the surface of this ellipsoid, so the ellipsoid solution
turned out to be impractical as well. Now, an infinitely long ellipsoid is identical with an infinitely long  cylinder, therefore a cosine coil
system on the surface of a cylinder can also be used to produce an approximately uniform magnetic field inside the cylinder.
In this arrangement,
 the uniformity of the field increases with the length of the cylinder and with the number of the current loops. 

\begin{figure}[htbp]
\centering
\subfigure[top]{\includegraphics[width=0.7\textwidth]{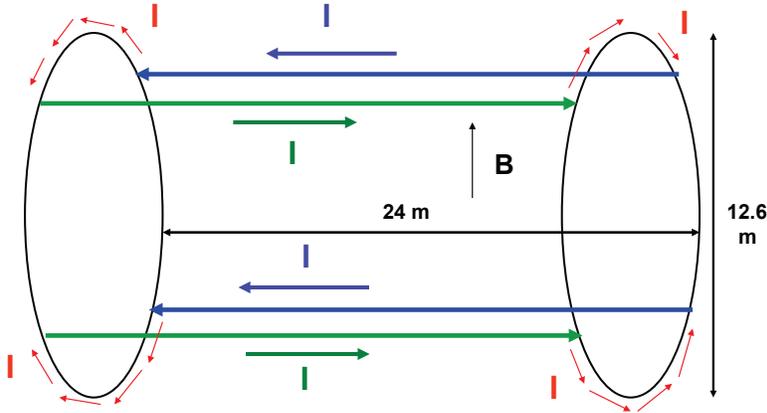}\label{FigLoopa}}\quad
\subfigure[bottom]{\includegraphics[width=0.7\textwidth]{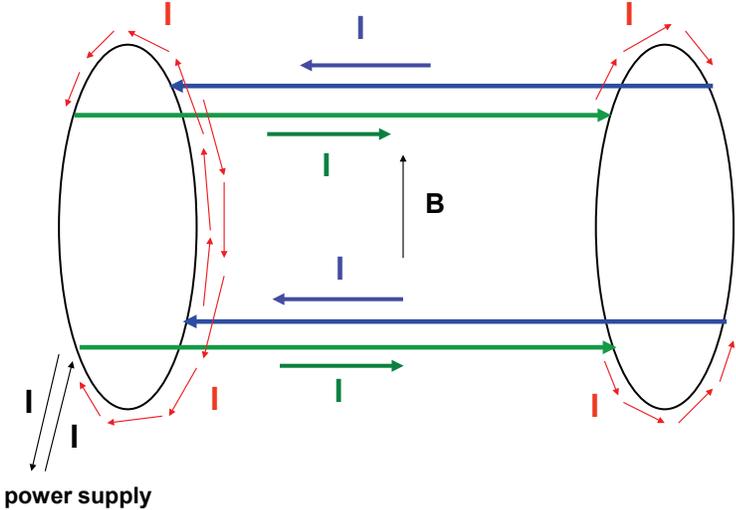}\label{FigLoopb}}         
\caption{Top: Two horizontal current loops for the compensation of the vertical earth magnetic field component.
Bottom: The two current loops united into one coil system.}
\label{FigLoop}
\end{figure}

\begin{figure}[htbp]
\centering
\includegraphics[width=0.75\textwidth]{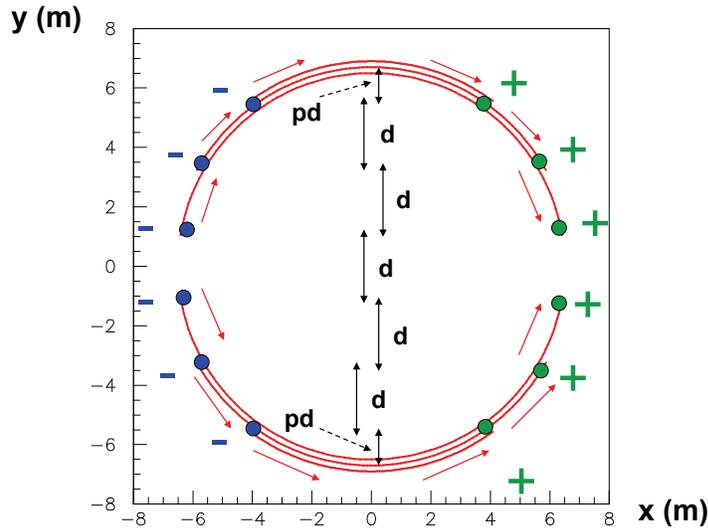}
\caption{Current flow at an endring of a simplified vertical compensation system with 6 current loops.}
\label{FigEndring}
\end{figure}

Accordingly, an air coil system on the surface of
a cylinder surrounding the KATRIN main spectromer tank turned out to be an optimal solution for the earth magnetic field compensation
inside the tank \cite{Gluck2004,Osipowicz,Gluck2008}. 
The length and radius of the cylinder was chosen to be 24 m and 6.3 m, respectively.
These dimensions were
constrained by the main spectrometer building, and they are identical  to the dimensions of the LFCS coils and
allow to construct both systems with a single mechanical support structure \cite{Reich2013}.
In order to compensate the vertical ($y$) component of the earth magnetic field
(43.6 $\mu$T), we have decided to use 16 current loops  with horizontal planes
(the blue lines in Fig. \ref{FigMainspec}), and for the horizontal transverse ($x$) 
earth field component (5 $\mu$T)
compensation we use 10 current loops with vertical planes (the red lines in Fig. \ref{FigMainspec}). 
Fig.  \ref{FigLoopa}  shows a current loop pair that provides a 
homogeneous vertical magnetic field at the center of the cylinder. One loop contains two linear current sections
(both of them parallel with the main spectrometer axis)  and, at the two endrings of the cylinder,  two arcs that connect
the linear sections, rendering the loop a closed current system. Fig.  \ref{FigLoopb}  
shows that the two closed loops are equivalent to one closed
current system that is  easier to realize practically. Similarly, the 16 current loops of the vertical  system 
and the 10 loops of the horizontal system are integrated into two independent closed current systems. Thus, the EMCS has
only two adjustable currents: 50 A  to produce a 43.6 $\mu$T field with the vertical system, and 9.1 A
to produce a 5 $\mu$T field with the horizontal system.

\begin{figure}[htbp]
\centering
\subfigure[horizontal field component]{\includegraphics[width=0.47\textwidth]{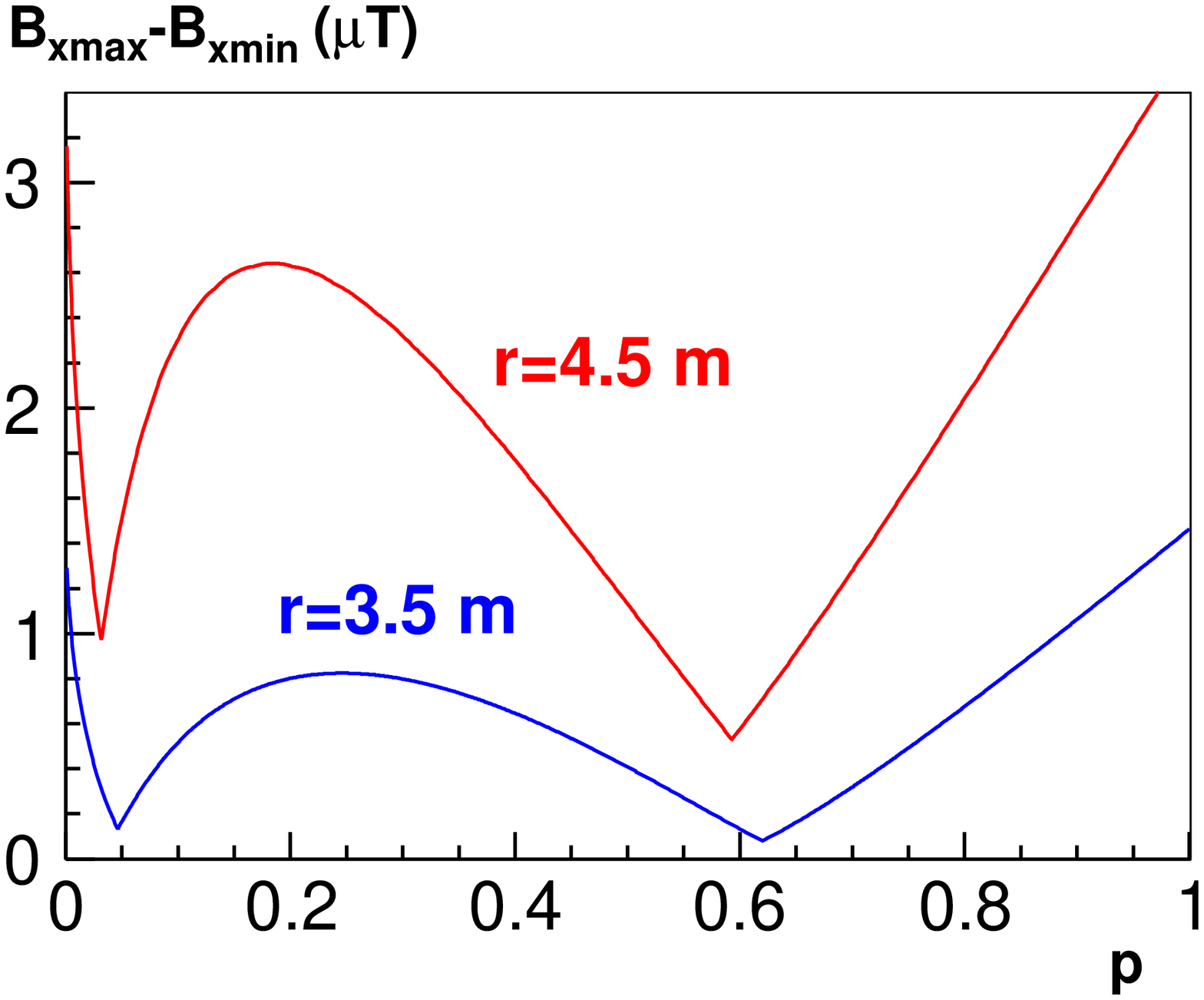}\label{Figpa}}\quad
\subfigure[vertical field component]{\includegraphics[width=0.47\textwidth]{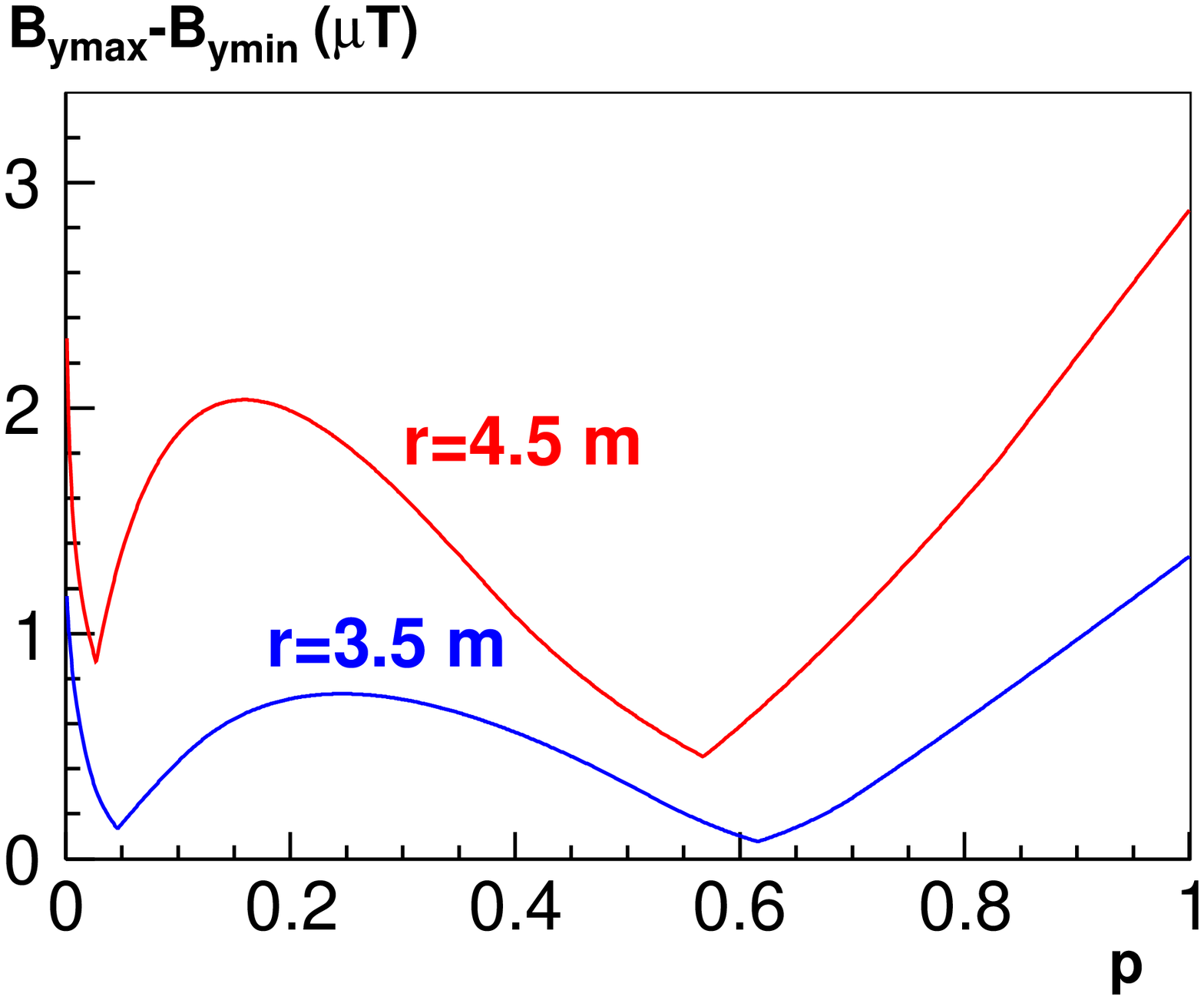}\label{Figpb}}         
\caption{Magnetic field inhomogeneity of the vertical part of the EMCS at the analyzing plane, as a function of
the end parameter $p$, for two different circles (with 3.5 m and 4.5 m radii).}
\label{Figp}
\end{figure}

\begin{figure}[htbp]
\centering
\subfigure[horizontal field component]{\includegraphics[width=0.47\textwidth]{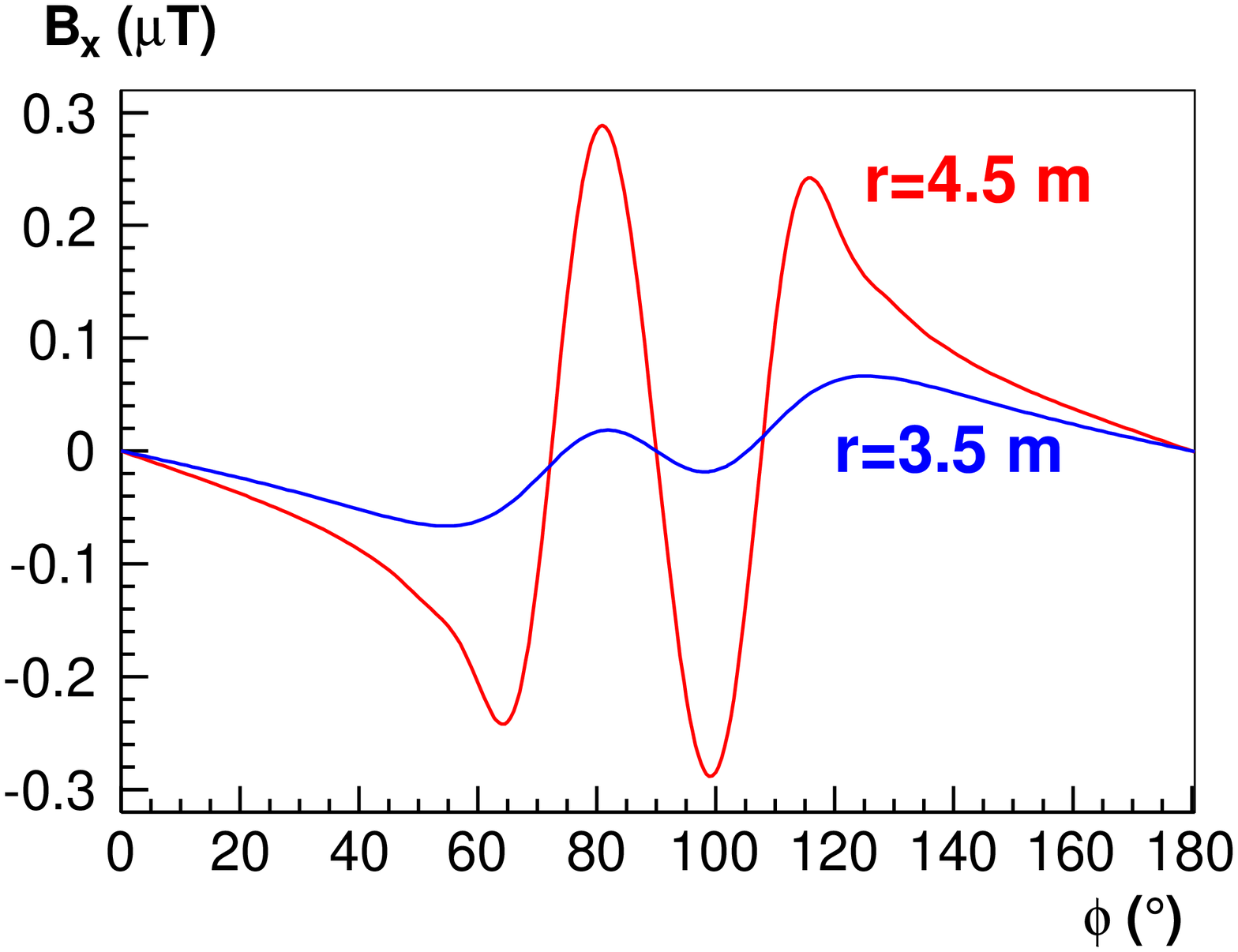}\label{Figya}}\quad
\subfigure[vertical field component]{\includegraphics[width=0.47\textwidth]{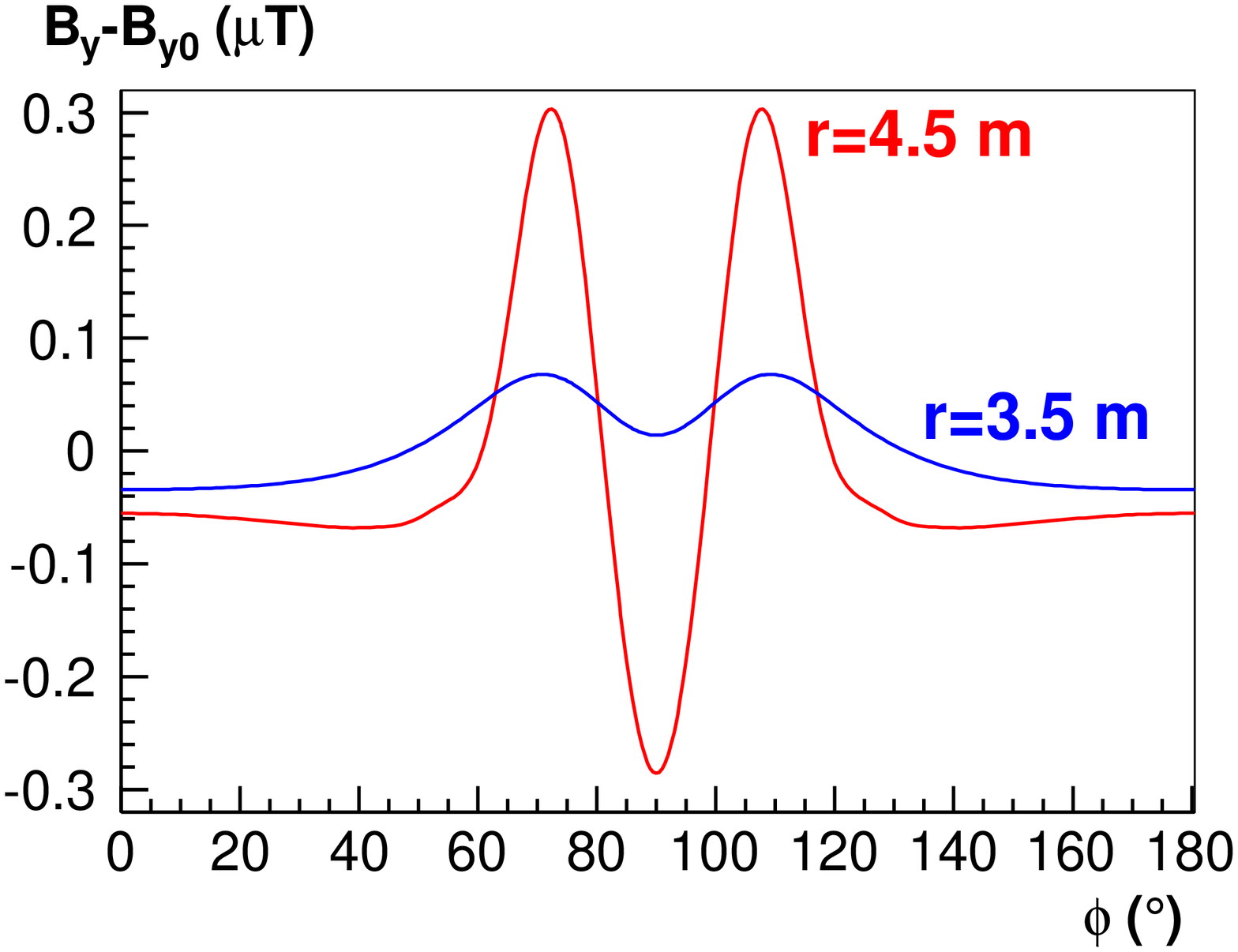}\label{Figyb}}         
\caption{The horizontal magnetic field component $B_x$ and the vertical field component difference
$B_y-B_{y0}$  of the vertical compensation system, 
as function of the azimuthal angle $\phi$ 
at two different circles (with 3.5 m and 4.5 m radii) in the analyzing plane. 
$\phi=0$ corresponds to the right-hand side point $x=6.3$ m, $y=0$  of Fig. \ref{FigEndring};
$\phi=90^\circ$ is for the top side point $x=0$, $y=6.3$ m.
The parameter $B_{y0}$=43.6 $\mu$T is the vertical field component of the vertical compensation system
at the center of the analyzing plane.}
\label{Figy}
\end{figure}

\begin{figure}[htbp]
\centering
\subfigure[horizontal EMCS]{\includegraphics[width=0.47\textwidth]{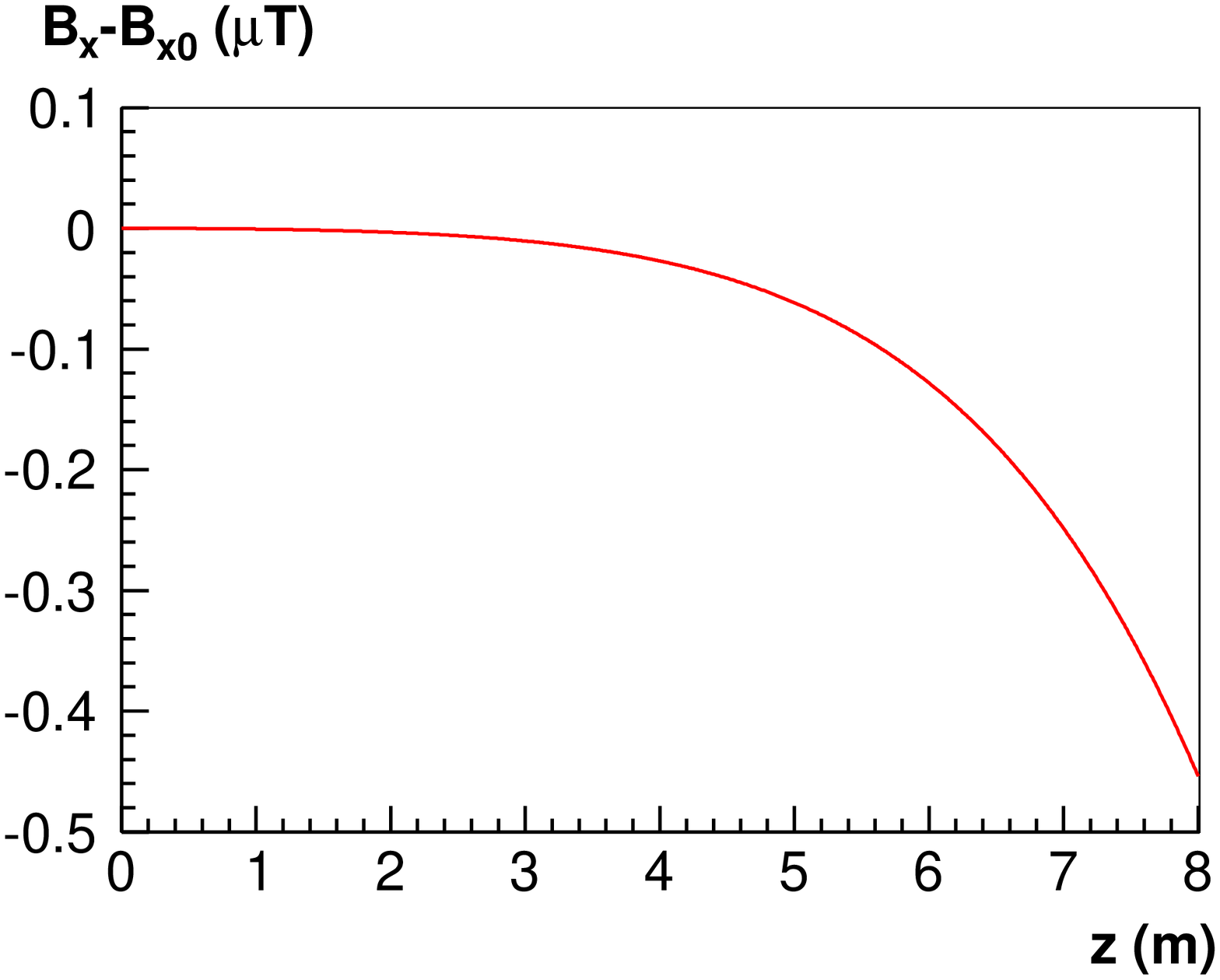}\label{Figza}}\quad
\subfigure[vertical EMCS]{\includegraphics[width=0.47\textwidth]{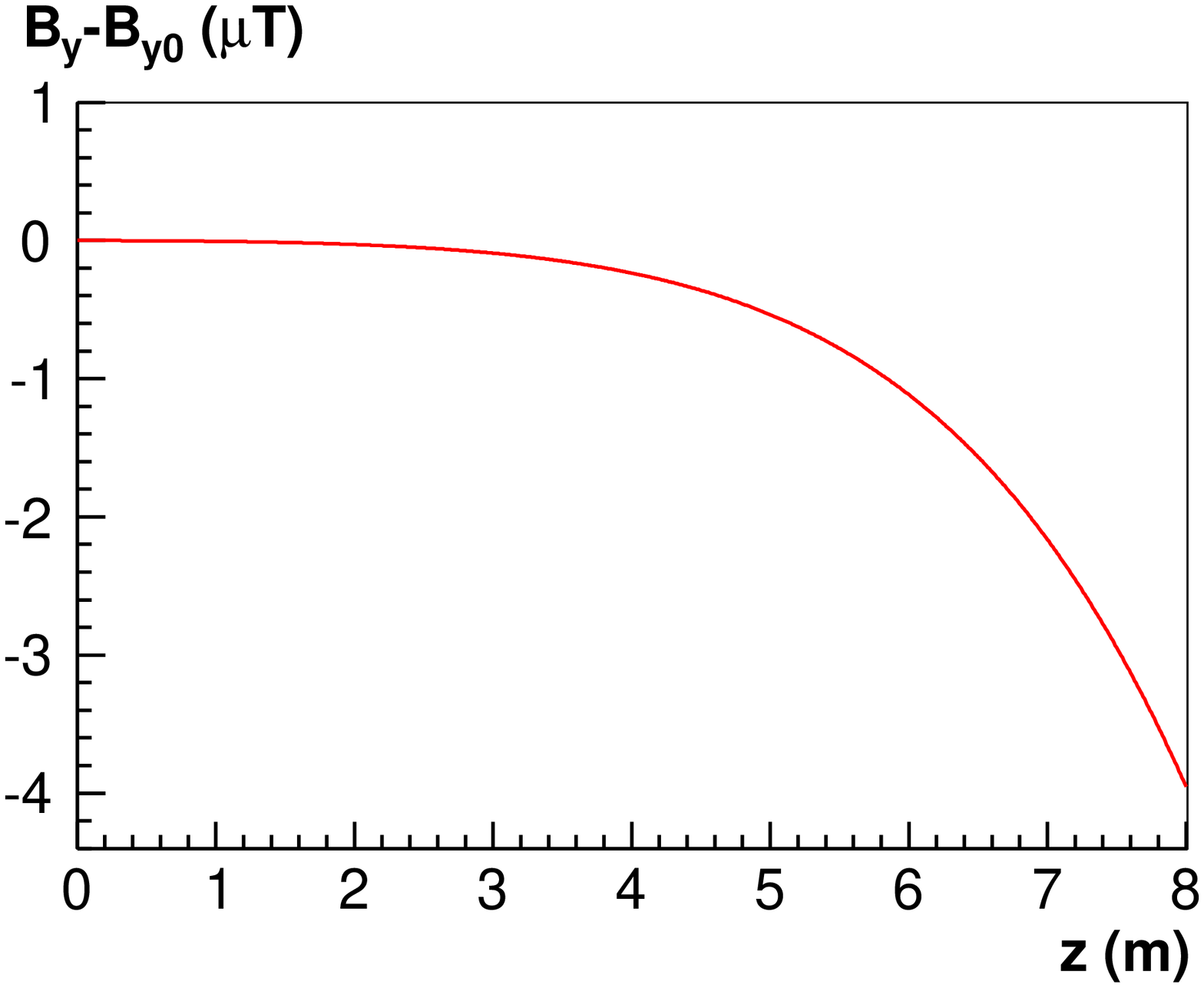}\label{Figzb}}         
\caption{On-axis magnetic field differences of the horizontal (left) and vertical (right) compensation systems, 
as function
of the distance $z$ from the analyzing plane. $B_{x0}$=5 $\mu$T is the horizontal $x$-component of 
the horizontal compensation system, 
$B_{y0}$=43.6 $\mu$T is the vertical $y$-component of 
the vertical compensation system, both
at the center of the analyzing plane.}
\label{Figz}
\end{figure}

Fig. \ref{FigEndring} illustrates the current arcs at one of the endrings in the special case of 6 loops. 
The positive signs mean that the current 
in the linear sections, which here are perpendicular to the page, flows in the direction inside the page, 
and the minus signs indicate
current flow direction outside the page (towards the reader). This figure shows  that the current loop planes are equidistant
(with distance $d$).
Identical current values of the equidistant loops correspond to an approximation of the
cosine current distribution and thus to uniform magnetic field.
An important design parameter of the EMCS is given by the so called end parameter $p$.
As can be seen in Fig. \ref{FigEndring}, the parameter $p\cdot d$ defines the distance of the outermost current loop to
the top or bottom of the ring elements, in case of equidistant arrangement of the current loops.
Accordingly, $p$ is a dimensionless free parameter with $0\le p \le 1$  \cite{Everett}.
Then, figs. \ref{Figpa} and  \ref{Figpb} show the inhomogeneity 
of the magnetic field components
$B_x$ and $B_y$ of the vertical ($y$-direction) compensation system at two circles with radii 3.5 m and 4.5 m,
as function of the end parameter $p$ (here the inhomogeneity has been defined as the difference of the
maximal and minimal field values on the circle).
One can see that the best field homogeneity is obtained for 
$p=0.6$.  Accordingly, we have chosen this value for both the vertical and the horizontal compensation systems.
Note that this optimal  value of $p$ for a cylindrical cosine coil system is different from the corresponding
optimal $p$ values of a spherical cosine coil system \cite{Everett}.

Figure \ref{Figy} illustrates the field inhomogeneity of the vertical compensation system at the analyzing plane
as a function of the azimuthal angle $\phi$ for two different radii.
Note that $\phi=0$ corresponds in Fig. \ref{FigEndring} to the point $x=6.3$ m, $y=0$.
One can see that the inhomogeneity increases with the distance from the spectrometer axis (at 
a radial position of $r=4.5$ m  from the axis
the inhomogeneity is several times larger than at $r=3.5$ m). The inhomogeneity of the field is maximal at the top and
bottom region of the coil system (at $\phi=90^\circ$ and $\phi=270^\circ$, respectively), 
where the deviation of the discrete coil
setup from the continuous $\cos\theta$  current distribution is maximal. Note that the vertical and the horizontal field components have
roughly the same  level of inhomogeneity.
The actual level of inhomogeneity of less than 0.3 $\mu$T in the analyzing plane, in comparison with the vertical and horizontal
components of the earth magnetic field, and in particular in relation to the absolute value of the guiding field of
0.35 mT, demonstrates the success of our optimization strategy in designing an effective EMCS.

The final important aspect of the EMCS design is the field behaviour along the longitudinal $z$-axis.
In this regard it is important to recall that the distorting effects of the earth magnetic field
have to be compensated mainly in the low field region $|z|<7$ m.
Figure \ref{Figz} shows that the field inhomogeneity of the horizontal and vertical compensation systems
increases with  the distance from the
analyzing plane (which is also the center of both the vertical and the horizontal coil system). 
This increase is due to the finite length of the coil systems and due
to the field disturbance from the circular current segments of the endrings. The lower quality of the compensation systems 
near the endrings should  not cause any  problem, since the magnetic field in the regions $|z|>7$ m is
already much larger than the earth magnetic field (see Fig. \ref{FigMagfield}).

The EMCS is useful not only to compensate the earth magnetic field, but also offers the useful possibility
to shift the magnetic flux tube.
The vertical and horizontal parts of the EMCS allow  shifting  the flux tube in the vertical and the horizontal
direction by 0.5 m, with 75 A and  50 A current, respectively \cite{Gluck2008}. 
These flux tube shifts can be important in order to correct some
small transversal shifts of the flux tube in the transport system, and also for specific background 
investigations and optimizations.

\section{Conclusion}
\label{SecConclusion}

The KATRIN experiment will determine the absolute neutrino mass scale down to 200 meV (90 \% CL)
by measuring the integral electron energy spectrum close to
the endpoint of molecular tritium beta decay. The $\beta$-electrons are guided from the source to the detector by 
magnetic fields, typically reaching values in the few T range in the source and transport system and being
created by many superconducting coils.
The energy filtering of the electrons takes place inside the large volume main spectrometer, which is at
high negative potential (around -18.6 kV).
In order to convert the transversal energy of the electrons into longitudinal energy by the inverse magnetic mirror effect and
thus to improve significantly the efficiency of the energy  filtering, the magnetic field strength in
the main spectrometer must reach very low values below 0.5 mT. 
The stray field of the superconducting coils alone is not sufficient to obtain the minimal 0.3 mT field that is needed to
constrain the magnetic flux tube to the geometry of the main spectrometer vessel. 
Moreover, the earth magnetic field disturbs significantly
this central low magnetic field region where the energy analysis takes place.
The task of the KATRIN large-volume air coil system described in this paper is to fine-tune and compensate these
fields.

The LFCS (Low Field Correction System)
part of the air coil system consists of 14 coils arranged coaxially with the main spectrometer vessel and the 
adjacent superconducting coils. With its help it is possible to set the magnetic field inside the main spectrometer from zero up
to 1 mT. The homogeneity of the field in the analyzing plane can also be improved considerably.
In addition, the asymmetric field of the superconducting coils 
can be compensated, thus making  the field more symmetric
relative to the $z=0$ analyzing plane. Even more importantly, with the LFCS one can fine-tune the magnetic field shape, adjusting it
to the electric potential, so that the 
adiabatic transmission condition is fulfilled. Thus it is much easier to evaluate accurately the transmission function 
of the MAC-E filter.
The precise knowledge of this function is an essential pre-requisite for a precision scanning of the
integral energy spectrum. To fulfill the transmission condition, two different possibilities have been worked out:
a magnetic field with a global minimum in the analyzing plane, and a field with a local maximum there but 
with two local minima a few meters away.
The second option has better theoretical properties: 
an easier fulfillment of the transmission condition and better homogeneity in the analyzing plane.
In order to find the optimal LFCS current values corresponding to these field alternatives, 
we have used a relatively simple and fast mathematical optimization method, 
based on a composite objective function with multiple objectives.

The second part of the air coil system is the EMCS whose task is to compensate those components of the  earth magnetic
field which are perpendicular to the spectrometer axis. It consists of two cosine coil systems: one of them compensates the
vertical earth magnetic field component (43.6 $\mu T$), the other one compensates the horizontal transversal earth field component 
(9.1 $\mu T$). In the analyzing plane of the main spectrometer, both the vertical and the horizontal components can be compensated
with 0.3 $\mu$T maximal inaccuracy, which is fully sufficient for high-precision $\beta$-spectroscopy.

The air coil system was constructed in 2009-2010.
Details about its mechanical and electrical layout, and about the commissioning field measurements and comparisons with
simulations will be presented elsewhere \cite{Reich2013}.

The KATRIN large volume air coil sytem will be an important experimental component for the main spectrometer commissioning
measurements, which will start in the first half of 2013.  
The purpose of these measurements is to examine and reduce the background,  and 
to investigate the electric, magnetic and electron transmission properties of the main spectrometer.
With the help of the LFCS and EMCS, one can set magnetic fields inside the main spectrometer 
in a highly versatile manner by adjusting both the overall field strength as well as the field shape.
Presumably, the background and transmission properties of the main spectrometer depend strongly on the LFCS and EMCS
currents, and we expect to find current values that result in a rather small background rate and well understood transmission function,
in order to obtain optimal conditions for the KATRIN neutrino mass measurements.

\section*{Acknowledgements}

This research was supported by the Helmholtz Association (HGF), the German Federal Ministry of Education and
 Research (BMBF), through grants 05A08VK2 and 05A11VK3, and the German Research Foundation (DFG)
within the framework of the Transregio project 'Neutrinos and Beyond', grant SFB/TR27. B.L., S.M. and N.W. thank the Karlsruhe
House of Young Scientists (KHYS) for supporting part of this study. We would like to thank Tudor Cristea-Platon
and Nils Stallkamp for their useful participation in the mathematical optimization computations,
and Michaela Meloni for her helping us to create  some of the figures.

\appendix

\section{Field simulations}
\label{appendix-fieldsim}

 In this work various field simulation codes have been used for the air coil design. The Part\-Opt code \cite{PartOpt}
uses elliptic integrals for magnetic field calculations of axisymmetric coils. 
In addition, the zonal harmonic expansion method was  employed \cite{Gluck2011b,Gluck2012a}.
The latter method can be 100-1000 times faster than the more
widely known elliptic integral method and  is more general than the similar radial series expansion.
It features not only high computational speed
but also high accuracy, which makes the method appropriate especially 
for trajectory calculations of charged particles. 

We could not use elliptic integrals or the zonal harmonic expansion to simulate the EMCS
since it is not axisymmetric.
Instead, the magnetic field of the linear current sections was computed by integrated Biot-Savart formulas \cite{Leiber}.
The arcs at the endrings were approximated by many short linear current segments.

In order to compute the adiabatic longitudinal energy, transmission energy and the analyzing points, we
also performed electric potential calculations.
 For this purpose, the boundary element method (BEM) was applied \cite{HawkesKasper,Corona}.
With BEM, one has to discretize only 
the two-dimensional surface of the electrodes, and not the whole three-dimensional space of the electrode system, as  is the
case when using  the finite difference and finite element methods \cite{HawkesKasper}. 
BEM is especially advantageous for electrodes exhibiting small-scale structures 
within large volumes, like the KATRIN wire electrode system \cite{Valerius}.
Inside the flux tube, the electric potential of the main spectrometer wire electrode system
is approximately axisymmetric, and with the knowledge of the charge densities from the BEM calculations
it is possible to use the zonal harmonic expansion method also
for the electric potential computations \cite{Gluck2011a,Gluck2012a}.

The field calculation C codes, written by one of us (F. G.), have been rewritten into C++ code
\cite{Corona,Leiber} and included
into the KASSIOPEIA package that is now the standard simulation framework of the KATRIN experiment
\cite{Kassiopeia,Furse}.

\section{LFCS current calculation by mathematical optimization}
\label{appendix-mathoptim}

Optimization problems naturally arise  in many different disciplines, like statistics, engineering, management, empirical sciences etc.
In mathematical (numerical) optimization \cite{NocedalWright,Bhatti,Snyman},
first one has to formulate  the problem. This is achieved by defining the design (optimization, decision)
 variables and the objective  (goal or cost) function
that has to be optimized (usually with some constraints on the design variables).  Then the optimal values of the
design variables leading to a  minimum of the objective function  
have to be found by applying some appropriate minimization technique. 
For more advanced problems, one has typically several different goals and
several requirements to be fulfilled simultaneously.
In this case one uses multiobjective (vector) optimization, with several objectives
to be optimized.
One of the possibilities to formulate this kind of optimization problem is by introducing a composite objective
function $F$ as the weighted sum of the objectives $O_k$:
\begin{equation}
\label{EqMultiobjective}
F=\sum_{k=1}^N w_k O_k .
\end{equation}
The weights $w_k$ have to be used so that the best result for the problem is obtained. The most important objectives and those with smaller
scalings  need  larger weight factors, 
so that these objectives should   decrease  significantly during the optimization procedure.

In our case, the design variables are the 14 LFCS  currents.
The magnetic field with the optimal current values has to fulfill several different requirements,
therefore we have adopted the multiobjective optimization procedure with composite objective function.
For the simulations yielding  the results of Sec. \ref{SecLFCS} we have used  $N=3$ objectives.

Our first objective was the squared deviation of the magnetic field value at the main spectrometer center from an input value:
$O_1=(B_0-B_{\rm input})^2$,
where $B_0=B(z=0,r=0)$; in our work  we have used  a value of $B_{\rm input}=0.35$ mT.

Our next goal was  to find a configuration where the magnetic field and the field lines 
are approximately perpendicular to the z=0 mirror plane.
In this case one can expect that the analyzing points are very close to this plane.
Therefore, for the second objective
we have defined an ensemble of
$n=10$ points at the $z=0$ mirror plane with $r_p=0.43\, p$ radius values (in meters)
$(p=1,\ldots,n)$, and we have computed the radial magnetic field components $B_r(p)$ at these points.
The second objective $O_2$ was then defined as the maximum
of the $|B_r(p)|$ values.

As for axisymmetric fields the radial component on the axis ($r=0)$ is always zero,
we have used also
a third objective. The goal here was to  have a magnetic field with extremum values in the $z=0$ plane.
 For this purpose, we have defined the set of
11 points with $r_p=0.43\, p$, $(p=0,\ldots,n)$, and we have computed there the
axial gradient field components $\partial_z B(p)$.  Then, the third objective
$O_3$  was defined as  the maximum of the $|\partial_z B(p)|$ values.
 For the computation of the axial gradients we used numerical differentiation:
$\partial_z B(p)\approx [B(z=\varepsilon,r=r_p)-B(z=-\varepsilon,r=r_p)]/(2\varepsilon)$, with $\varepsilon=0.1$ mm.

In the next step, the composite objective function is the weighted sum of these 3 objectives (see Eq. \ref{EqMultiobjective} for $N=3$).
We have chosen to use the empirical weight factors  $w_1=1$, $w_2=w_3=10$.

The points where we have computed the magnetic field values are fixed (they are independent of the optimization procedure),
therefore we have been able to reduce significantly the required computation time 
by calculating, in the beginning, for all points the magnetic field
contributions $b_j$ of  the LFCS coil $j$ with 1 A current. Then, during the optimization, the field can be computed rapidly
 as the linear superposition
$B=B_{\rm sc}+\sum_{j=1}^{14} b_j I_j$, where  $B_{\rm sc}$ denotes the field due to the superconducting coils 
and the horizontal earth magnetic field.

To minimize the objective function $F$, starting at some point in the 14 dimensional current space,
we have used the Nelder-Mead downhill simplex method (\cite{NelderMead}, and  \cite{Press}, sec. 10.4).
This popular minimization method requires only the evaluation of functions, 
and not their derivatives. It is based on the notion of a simplex
that is a geometrical figure having $n+1$ points (vertices) in the n-dimensional 
design variable space (the simplex is the generalization of a triangle or tetrahedron 
for higher dimensions). In the beginning, a simplex is created near the starting point, and it is changed by various 
transformations (reflection, contraction, expansion, shrinkage) so that the average function value at the simplex vertices
continuously decreases, until the simplex attains a local minimum of $F$ 
where no further significant reduction of the function value is possible.

From the technical point of view, 
the LFCS currents are not allowed to exceed the upper limits presented in Table \ref{TabLFCS}.
In our work, we used the following limits: $I_{\rm min}=-100$ A,  $I_{\rm max}=0$ for coils 1-13, 
$I_{min}=0$,  $I_{max}=70$ A for coil 14. In order to include these limits as constraints into our optimization code,
we introduced the following variable transformation: 
$I_j=I_{{\rm min},j}+(I_{{\rm max},j}-I_{{\rm min},j})\cdot (1+\cos x_j)/2$. Using the variables $x_j$ for the function minimization, instead 
of the currents $I_j$, the constrained optimization is turned into the easier case of  unconstrained optimization
(the variables $x_j$ can have arbitrary values, while the currents are  constrained between their lower and upper limits).

As we have mentioned in Sec. \ref{SecLFCS}, the starting point for our mathematical optimization procedure
was the result of a first, rough optimization-by-eye operation. We tried our mathematical optimization process also
by arbitrary (randomly chosen) starting points. In that case, too, the minimization by the simplex method was able
to reduce significantly  the objective function value and  found some local minimum. Unfortunately, 
in this case the current values
of the neighbouring coils corresponding to these local minima featured rather large jumps, resulting in unnecessarily large currents
for some of the coils. Probably, one has to use some additional objectives (like the total electric power of the coils) to avoid 
these large current jumps. These on-going investigations, however, will not influence our conclusions presented above.

\begin{comment}

\begin{figure}[htbp]
\centering
\subfigure[left]{\includegraphics[width=0.47\textwidth]{FigFieldlines1.eps}\label{FigFieldlines1}}\quad
\subfigure[right]{\includegraphics[width=0.47\textwidth]{FigFieldlines2.eps}\label{FigFieldlines2}}         
\caption{Magnetic field lines of the 191 ${\rm Tcm}^2$ flux tube. {\it Left}: 0.35 mT at center, global minimum in AP ( analyzing plane);
{\it Right}: 0.35 mT at center, local maximum in AP and off-axis (and 2 local minima).}
\label{FigFieldlines}
\end{figure}

 Citations for Sec. 4 d,

Energy resolution formula

Explain the max. 51 deg angle in WGTS

\bibitem{Osipowicz2012a} A. Osipowicz et al.,
A mobile magnetic sensor unit for the KATRIN main
spectrometer,
J. Instrum. 7 (2012) T06002.
\href{http://dx.doi.org/10.1088/1748-0221/7/06/T06002}
{(hyperlink)}

\bibitem{Osipowicz2012b} A. Osipowicz et al.,
 A scheme for the determination of the magnetic field in the KATRIN
main spectrometer,
arXiv:1209.5184.
\href{http://arxiv.org/pdf/1209.5184.pdf}
{(hyperlink)}

\end{comment}

\end{document}